\documentclass{PoS}

\usepackage{amssymb,amsmath,amsfonts,amsthm,bbold}
\usepackage{mathrsfs,mathtools}
\usepackage{graphicx}

\usepackage{minibox} 

\usepackage{float}
\usepackage{caption}
\usepackage{subcaption}

\usepackage{bbm}

\newcommand{\ket}[1]{|#1\rangle}
\newcommand{\bra}[1]{\langle#1|}

\newcommand{\Am}{\ensuremath{\mathcal{A}}}
\newcommand{\LF}{\left(}
\newcommand{\RF}{\right)}
\newcommand{\LT}{\left[}
\newcommand{\RT}{\right]}
\newcommand{\Ym}{\ensuremath{\mathcal{Y}}}

\title{Modave lectures on bulk reconstruction in AdS/CFT}

\ShortTitle{Modave lectures on bulk reconstruction in AdS/CFT}

\author{\speaker{Tim De Jonckheere}\\
        *Theoretische Natuurkunde, Vrije Universiteit Brussel, and \\ International Solvay Institutes,
Pleinlaan 2, B-1050 Brussels, Belgium \\
        E-mail: \email{tim.de.jonckheere@vub.be}}

\abstract{These lecture notes are based on a series of lectures given at the XIII Modave summer school in mathematical physics. We review the construction due to Hamilton, Kabat, Lifschytz and Lowe for reconstructing local bulk operators from CFT operators in the context of AdS/CFT and show how to recover bulk correlation functions from this definition. Building on the work of these authors, it has been noted that the bulk displays quantum error correcting properties. We will discuss tensor network toy models to exemplify these remarkable features. We will discuss the role of gauge invariance and of diffeomorphism symmetry in the reconstruction of bulk operators. Lastly, we provide another method of bulk reconstruction specified to AdS$_3$/CFT$_2$ in which bulk operators create cross-cap states in the CFT.}

\FullConference{XIII Modave Summer School in Mathematical Physics-Modave2017\\
		10-16 September 2017\\
		Modave, Belgium}

\begin{document}

\section*{Note to the reader}
These notes are based on a series of 5 hours of lectures taught at the XIII Modave Summer School in Mathematical Physics. The main purpose of the notes is to familiarize graduate students with the notion of bulk reconstruction in the context of AdS/CFT. In no way do these notes cover the complete scope of the research field nowadays called bulk reconstruction, and the topics presented here are just a selection made by the author. For a further in depth study of this interesting subject, the author would like to refer to the bibliography included here and the bibliographies of the cited papers. 

\section{A brief history of bulk reconstruction}

Two decades after its discovery, the AdS/CFT duality~\cite{Maldacena:1997re,Witten:1998qj} has grown into a research topic on its own away from its string theory origins, and more broadly speaking has led to a new perspective on both gravitational and gauge theory physics. It postulates a duality between a conformal field theory in $d$ dimensions and a gravitational theory that leads to an asymptotically AdS spacetime in $d+1$ dimensions. The conformal field theory is visualized as living on the boundary of the gravitational spacetime. For that reason the AdS/CFT duality is also called holographic dictionary. A duality implies that the AdS Hilbert space of physical states should be isomorphic to the CFT one and that the algebra of operators $\Am(\hat{O})$ of the two theories should be isomorphic as well. Local bulk operators $\hat{\phi}$ should therefore be expressable solely in terms of CFT operators $\hat{O}$. The purpose of bulk reconstruction is to clarify that map between bulk and boundary operators. Suppose the boundary theory is known, which means all of its operators and correlation functions are known, then the theory is said to be completely `solved'. Knowing the map between bulk and boundary operators would therefore mean that also the gravitational theory has been fully `solved'.\\

The bulk reconstruction program has been initiated in the early days of AdS/CFT duality~\cite{Balasubramanian:1999ri,Giddings:1999qu} where the CFT correlation functions were interpreted as gravitational S-matrix elements in a bubble of AdS space, and local bulk operators were linked to boundary operators via bulk-boundary propagators. A few years later, the study of bulk reconstruction was developed more rigourously by Hamilton, Kabat, Lifschytz and Lowe in a series of seminal papers~\cite{Hamilton:2005ju,Hamilton:2006fh,Hamilton:2006az,Hamilton:2007wj}. The authors explicitly solved the bulk equations of motion of a free scalar with sourceless boundary conditions\footnote{We will review in the next sections what is meant with this statement.} and expressed the solution by inverting a mode expansion as an integral of local CFT operators smeared over an appropriate boundary region with a kernel called the \textit{smearing function}.\\

The HKLL\footnote{for Hamilton, Kabat, Lifschytz and Lowe.} smearing function procedure that reproduces a bulk operator as a set of non-locally smeared CFT operators 
has several shortcomings. A big part of these lectures will consist of a derivation and discussion of the HKLL reconstruction program and its problems. Finding an answer to these shortcomings has sparked a lot of interest in the bulk-reconstruction research area and will be a central subject of these lectures. It was pointed out in~\cite{Almheiri:2014lwa} that in fact the HKLL reconstruction scheme leads to a paradox once multiple possible reconstruction regions are considered. To get around the paradox, the authors of~\cite{Almheiri:2014lwa} conjectured we should view the AdS/CFT dictionary as a quantum error correcting code (QECC). For some time the AdS/CFT community has been unfamiliar with quantum error correcting codes which are a standard subject in quantum information theory. The link between AdS/CFT and quantum information theory has in fact been pointed out in many recent works~\cite{Swingle:2009bg,Swingle:2012wq,Qi:2013caa}. The underlying idea in much of this work is that the bulk geometrizes the entanglement structure of the boundary state and most examples are based on a tensor network construction of the boundary state. In these lectures we will very shortly review what tensor networks are and how they are linked to holography. We will then proceed by discussing the paradox put forward in~\cite{Almheiri:2014lwa} and discuss the quantum error correcting properties of AdS/CFT. To do so we will make use of a particular tensor network, put forward in~\cite{Pastawski:2015qua}. This code consists of very particular tensors such that the resulting network acquires part of the AdS isometry group. It can be shown for such networks that the entanglement entropy can be computed by the area of an extremal surface in the network, which is an essential feature of AdS/CFT. Secondly, this network precisely has the conjectured quantum error correcting properties.\\

The essential property of QEC in the context of bulk reconstruction is the fact that operators  that are distinct on the boundary Hilbert space cannot be distinguished on the effective field theory subspace and therefore have the same action on the latter. An important aspect that has been completely left out of the story of reconstruction is the built in gauge symmetry that is present in holographic theories. In fact, as it turns out, gauge symmetry displays the same essential feature as a QEC~\cite{Mintun:2015qda,deBoer:2016yiz,Freivogel:2016zsb}. We will study the very simple toy model of~\cite{Mintun:2015qda} to exemplify these features. Although the picture of the effect of gauge symmetry on the reconstruction has not completely been worked out yet, it would provide a much more natural explanation for the apparent paradox with HKLL than an artificial reinterpretation of the holographic dictionary as a quantum error correcting code. We will illustrate how gauge invariance can lead to the QEC properties of the bulk. Another important aspect is the diffeomorphism symmetry. Well defined bulk operators are necessarily diffeomorphism invariant. This invokes the need for gravitational dressing~\cite{Dirac:1955uv,Donnelly:2015hta} and we will show that the dressing turns local bulk operators effectively into non-local ones. The dressing can be combed~\cite{Donnelly:2015taa} such that the operator only has support on a particular boundary region. The gravitational dressing in a large $N$ holographic theory is $1/N^2$ suppressed. Local bulk operators that only differ in their gravitational dressing and the associated boundary region of support are therefore equivalent at the classical level. This connects to the paradox of~\cite{Almheiri:2014lwa}.\\

Finally, in $AdS_3/CFT_2$ one does not need to resort to HKLL in order to reconstruct bulk operators. In this setting, restricting to pure $d=2+1$ gravity without matter, it can be argued that a bulk operator at a point creates a state in the CFT called a cross-cap state~\cite{Miyaji:2015fia,Nakayama:2015mva,Nakayama:2016xvw,Verlinde:2015qfa,Lewkowycz:2016ukf}. The cross-cap states actually change the topology of the underlying complex manifold on which the CFT is defined and create a non-orientable surface which is obtained by $\mathbbm{Z}_2$ identifications on the original manifold. Bulk operators would therefore always act very non-local on the boundary according to this proposal. Nevertheless, this mapping between bulk operators and cross-cap operators also has a couple of clear advantages. It is possible to define bulk operators in such a way that they coincide with HKLL at leading order in $1/N$, but are gravitationally dressed at the non-perturbative level and the dressing can be derived. They also provide operators which are independent of the background geometry and it can be shown that standard bulk correlators are reproduced from cross-cap correlation functions. We will review these features in the lecture notes here.\\

The lectures will focus on the reconstruction of bulk operators from the CFT, and to explain a duality between gravitational and gauge theories, explaining the isomorphism between the operator algebras is enough, but for the gravity theory to really be holographic, one expects the geometry itself to be emergent from the CFT. A remaining question then is what the governing principle is according to which the spacetime is emergent. The modern point of view is that the spacetime might emerge from the entanglement in the boundary~\cite{Ryu:2006bv,VanRaamsdonk:2009ar}. Even before the discovery of AdS/CFT, Hawking~\cite{Hawking:1974sw} and Bekenstein~\cite{Bekenstein:1973ur} connected gravity to thermodynamics. They discovered that the area of a black hole is proportional to its thermal entropy. In 2006, Ryu \& Takayanagi~\cite{Ryu:2006bv} formulated a similar idea in the context of AdS/CFT: the entanglement entropy of a subsystem $A$ on the boundary is proportional to the area of a surface in the bulk that is homologous to the boundary subsystem and that ends on the boundary of that subsystem $\partial A$
\begin{equation}
S_A \equiv - \text{Tr} \LF \rho_A\log\rho_A\RF=\frac{\Am}{4G_N}.
\end{equation} 
Since a gravitational spacetime can be sliced into surfaces, this leads to the idea that gravity could be emergent from the CFT by ``geometrizing" its entanglement structure and has sparked a tremendous amount of literature over the past ten years.\\

The entanglement entropy has been studied in various kinds of static and dynamic backgrounds and more general surfaces have been matched to entanglement entropy related quantities like differential entropy~\cite{Balasubramanian:2013lsa,Headrick:2014eia} or entwinement~\cite{Balasubramanian:2014sra,Balasubramanian:2016xho}. In fact, the study of entanglement entropy and related quantities in order to study the emergence of geometry has grown into such a broad subject that it would take at least a series of lectures on its own. For that reason, we choose not to elaborate on this aspect of reconstruction.\\

We should also emphasize that the lecture notes contain by no means a discussion on every possible procedure of reconstructing the bulk. A few examples of other methods that will not be discussed here, but have been proposed in the literature are reconstruction with use of the modular Hamiltonian~\cite{Faulkner:2017vdd} and the light-cone cut approach to reconstruction~\cite{Engelhardt:2016wgb}.\\

The outline of the lectures is as follows: we will review the basics of AdS/CFT and discuss a scalar field in AdS. Readers familiar with the basics can safely skip this section and go to section~\ref{sec:HKLL} where the HKLL formalism for reconstructing local bulk operators is presented. Then in section~\ref{sec:QEC} we derive the paradox associated with HKLL and discuss the resolution through quantum error correction. Because QEC is necessarily tied to information theory which is often studied through tensor networks, we also give a very brief and qualitative discussion of tensor networks. Section~\ref{sec:gauginv} relates the QEC properties of holography to the built-in gauge invariance of holographic theories and contains a derivation of gravitational dressing in the particular case where the dressing operator is a gravitational Wilson line. Finally, in section~\ref{sec:crcap} we construct bulk operators as operators that create cross-cap states in the boundary CFT. Both in section~\ref{sec:HKLL} and \ref{sec:crcap} we moreover show that the bulk-boundary correlator is correctly reproduced. Our concluding remarks are presented in section~\ref{sec:conclusion}.

\section{An AdS/CFT reminder}
The AdS/CFT duality originates from string theory. It was argued~\cite{Maldacena:1997re,Polchinski:1995mt} that $D$-branes on which open strings end, when $N$ of them are stacked together and an $SU(N)$ gauge group emerges, are also the sources of closed string excitations. Open strings describe the $SU(N)$ gauge sector while closed string excitations give rise to gravitation. The equivalence of branes as gravitating objects and as sources of $SU(N)$ fields leads to a duality between gravity and gauge theory. Consider the example of $N$ stacked $D3$ branes. At low energies and in the limit $N\rightarrow + \infty$ the open string description reduces to a $SU(N)$, $\mathcal{N}=4$ SYM theory in $d=3+1$ dimensions. The same system in the closed string description at low energies gives rise to an $AdS_5\times S^5$ spacetime. The $S^5$ is decoupled from $AdS_5$ and represents only the internal $R$-symmetry of the $SU(N)$ gauge theory. Moreover, the $d=3+1$ $\mathcal{N}=4$ SYM theory is conformally invariant, so we have an $AdS_5/CFT_4$ duality. Because the gravitational spacetime is of $1$ dimension higher than the gauge theory spacetime, we call the gravitational spacetime the bulk. Since the asymptotic boundary of $d+1$-dimensional AdS is $d$-dimensional Minkowski spacetime, we can define the CFT on the asymptotic boundary of the bulk spacetime. The boundary is only defined asymptotically, so an IR regulator needs to be defined in the bulk. This IR bulk cut-off is dual to the UV-cutoff of the CFT, which is why the AdS/CFT duality is also called UV/IR duality.\\
Another useful example is that of $N_1$ $D1$-branes and $N_5$ $D5$-branes. This system reduces to an $AdS_3\times S^3\times T^4$ spacetime on the one hand and a $d=1+1$ orbifold CFT with target space \newline $\LF T^4\RF^{N_1N_5}/S_{N_1N_5}$ on the other hand. Although we will not use the string theory details, it is good to know there exists a string theory realization of an $AdS_3/CFT_2$ duality as well. This is useful because we will very often in this course specialize to the duality in $d=2$. The examples so far are dualities involving AdS, but the conjecture can be generalized to asymptotically AdS spacetimes. An example of this is a black hole spacetime with asymptotic AdS boundary conditions. Such a spacetime is thought to be dual to a thermal state of the corresponding CFT.\\

The string theory details are not needed for a lot of the computations in AdS/CFT. For example, for computing correlation functions it is enough to know the CFT fusion coefficients. The correlation functions are dual to gravitational amplitudes. Similarly, it is not necessary to know the precise string theory embedding to be able to reconstruct bulk local operators from the dual CFT. To set the scene for the HKLL program, we start with a discussion of various useful coordinate sets in AdS, and mention a couple of relevant CFT facts.

\subsection{AdS}
\subsubsection{Coordinate systems}
Anti-de Sitter space in $d+1$ dimensions is specified as a hyperbolic surface in a $d+2$ dimensional Minkowski space of metric $ds^2 = -dX_0^2 - dX_1^2 + dX_2^2 + \ldots + dX_{d+1}^2$. The hyperboloid that specifies AdS is
\begin{equation}
-X_0^2 - X_1^2 + X_2^2 + \ldots X_{d+1}^2 =- R^2
\end{equation}
with $R$ the AdS radius which will be set to unity most of the times. A set of useful coordinates that cover the full AdS spacetime are the global coordinates, which when $d=2$ are
\begin{equation}
\left\{ \begin{matrix}
X_0 = \sqrt{r^2+R^2} \cos (t/R),\\
X_1 = \sqrt{r^2+R^2} \sin(t/R),\\
X_2 =  r \sin\varphi,\\
X_3 =  r\cos\varphi.
\end{matrix}
\right.
\end{equation}  
The anti de Sitter space looks like a cylinder in these coordinates with $t$ the time running upwards, $\varphi$ the angular coordinate and $r$ the radial coordinate. The metric in these coordinates is
\begin{equation}
ds^2 = -(\frac{r^2}{R^2}+1)dt^2 + \frac{dr^2}{\frac{r^2}{R^2}+1} + r^2d\varphi^2
\end{equation} 
or with $r= R\tan\rho$
\begin{equation}
ds^2 = \frac{R^2}{\cos^2\rho} \LF - dt^2  + d\rho^2 + \sin^2\rho d\varphi^2\RF
\label{eq:globalcoord}
\end{equation}
 Another famous set of coordinates are the Poincar\'e coordinates
\begin{equation}
\left\{ \begin{matrix}
X_0 = \frac{R^2 y}{2} \LF 1+ \frac{1}{R^4 y^2} \LF R^2 +x^2-t^2\RF\RF,\\
X_1 = \frac{t}{Ry},\\
X_2 = \frac{x}{Ry} ,\\
X_3 = \frac{R^2y}{2} \LF 1- \frac{1}{y^2 R^4}\LF R^2-x^2+t^2\RF\RF.
\end{matrix}
\right.
\end{equation}
The metric in these coordinates is simply $ds^2 = \frac{1}{y^2} \LF -dt^2 +dy^2+dx^2\RF$. Poincar\'e coordinates only cover a patch of AdS.\\

Lastly, a set of coordinates that is particularly useful for bulk reconstruction are the Rindler coordinates with associated Rindler metric (see e.g., \cite{Parikh:2012kg}),
\begin{equation}
ds^2 = -\LF \frac{\xi}{R}\RF^2 d\eta^2+ \frac{d\xi^2}{1+ \frac{\xi^2}{R^2}} + \LF 1+ \frac{\xi^2}{R^2}\RF d\chi^2.
\end{equation}
The embedding coordinates are now expressed as 
\begin{equation}
\left\{ 
\begin{matrix}
X^0 = \xi \sinh\LF \frac{\eta}{R}\RF,\\
X^1 = \sqrt{R^2+\xi^2}\cosh\LF \frac{\chi}{R}\RF,\\
X^2 = \sqrt{R^2+\xi^2}\sinh\LF \frac{\chi}{R}\RF,\\
X^3 = \xi \cosh\LF \frac{\eta}{R}\RF.
\end{matrix}
\right.
\end{equation}
The Rindler patch covers $(X^3)^2-(X^0)^2\geq 0$ which is half of the AdS space. There is a Rindler horizon at $X^0=\pm X^3$ when $\eta\rightarrow \pm \infty$.
\subsubsection{Scalar field in AdS}
Consider a free scalar field on AdS. Its equations of motion are
\begin{equation}
\LF \Box - m^2\RF \phi = 0.
\end{equation}
This can easily be solved in terms of a mode expansion which is what we will do later on in these notes. For later purpose we will parametrize $m^2 = \Delta (\Delta -d)$.\\

The goal of these lectures is not just to solve for the field but also to compute amplitudes on AdS. A simple example is the scalar amplitude $\langle \phi(\rho_1,\varphi_1,t_1) \phi(\rho_2,\varphi_2,t_2)\rangle$ which is just the Green's function on AdS.  Because of the $O(d,2)$ isometry group of AdS, the Green's function can only depend on an invariant distance function\footnote{The expression for the invariant distance can be obtained by starting from the distance in Minkowski space in embedding coordinates and pulling back to the AdS hyperboloid expressed in global coordinates.}  $\sigma$ between the two insertion points of the scalar field with
\begin{align}
\sigma &= \frac{ \cos(t_1-t_2)-\sin\rho_1\sin\rho_2\cos(\varphi_1-\varphi_2)}{\cos\rho_1\cos\rho_2}.
\end{align}
Going to Euclidean signature, writing out the EOM on AdS and performing a pull-back to $\sigma$, one arrives at the ODE~\cite{Hamilton:2006az}
\begin{equation}
\LF \sigma^2 -1 \RF G_E^{''} + \LF d+1\RF \sigma G_E^{'} - \Delta \LF \Delta -d \RF G_E =0
\end{equation}
but with a $\delta$-function source at $\sigma=1$. When $d=2$ we can easily solve the equation. By redefining $z=\sigma^2$ this is turned into the hypergeometric equation, outside the $z=1$ point (such that the delta function is neglected).
\begin{equation}
z(1-z)\frac{d^2G_E}{dz^2} + \LF \frac{1}{2}-2z\RF \frac{dG_E}{dz} - \frac{1}{4}\Delta(2-\Delta)G_E = 0
\end{equation}
with solution $G_E(z) = z^{-\frac{\Delta}{2}}\left._2 F_1\right.\LF \frac{\Delta}{2},\frac{1+\Delta}{2},\Delta,\frac{1}{z}\RF$. We have picked the solution to be singular at $z=1$ to recover the $\delta$-function. We substitute back $z=\sigma^2$, define the geodesic distance $\xi = \ln \LF \sigma + \sqrt{\sigma^2-1}\RF$ and evaluate the hypergeometric function for these particular arguments to obtain
\begin{equation}
G_E(\xi) = \frac{e^{-\Delta \xi}}{e^{-2\xi}-1}.
\end{equation}
In the limit where one of the fields approaches the boundary, we retrieve the Euclidean \textit{bulk-boundary} propagator in AdS
\begin{equation}
G_{E,b\partial}\LF \rho,t_1,\varphi_1;t_2,\varphi_2\RF = \LF \frac{ \cos\rho}{\cosh(t_1-t_2) - \sin\rho \cos(\varphi_1-\varphi_2)}\RF^{\Delta}.\label{eq:bbprop}
\end{equation}

\subsection{CFT}
A \textit{conformal field theory} (CFT) is a quantum field theory that is invariant under conformal transformations, i.e.  $x'^{\mu} = x'^{\mu}(x)$ such that $g'_{\rho\sigma}(x') \frac{\partial x^{'\rho}}{\partial x^\mu} \frac{\partial x^{'\sigma}}{\partial x^\nu} = \Lambda(x) g_{\mu\nu}(x)$ with $\Lambda$ a local scale factor.\\

It is possible to formulate CFT's in any dimension $d$, but in $d=2$ the conformal symmetry is enhanced to a local conformal symmetry transformation with arbitrary (anti-) holomorphic symmetry generators, namely $ds^2 = dz'd\bar{z}^{'}$ with 
\begin{equation}
\left\{ 
\begin{matrix}
z' = z + \epsilon(z),\\
\bar{z}' = \bar{z} + \bar{\epsilon}(\bar{z}).
\end{matrix}
\right.
\end{equation}
Because these infinitesimal transformations are just reparametrisations of the $z$ coordinate the generator of the conformal transformations is just the stress tensor. Because the holomorphic structure is preserved in $d=2$, the only non-trivial stress energy components are $T_{zz}= T(z)$ and $T_{\bar{z}\bar{z}}= \bar{T}(\bar{z})$. The stress tensor can be decomposed into a Laurent mode expansion
\begin{equation}
\left\{
\begin{matrix}
T(z) = \sum\limits_{n\in \mathbbm{Z}} L_n z^{-n-2},\\
\bar{T}(\bar{z}) = \sum\limits_{n\in\mathbbm{Z}} \bar{L}_n \bar{z}^{-n-2}.
\end{matrix}
\right.
\end{equation}
The $L_n,\bar{L}_n$ modes are called \textit{Virasoro} modes and satisfy the Virasoro algebra~\cite{DiFrancesco:1997nk}.
\begin{equation}
\LT L_n, L_m\RT = (n-m) L_{m+n} +\frac{c}{12}\LF n^3-n\RF \delta_{n+m,0}.
\end{equation}
Operators that are covariant under conformal transformations are called \textit{primary} operators. They transform as $O_\Delta (z,\bar{z}) = \LF \frac{\partial z'}{\partial z}\RF^h \LF \frac{\partial\bar{z}'}{\partial \bar{z}}\RF^{\bar{h}} O_\Delta \LF z(z'),\bar{z}(\bar{z}')\RF$ with $\Delta= h+\bar{h}$ the same $\Delta$ as $\Delta(\Delta-d)=m^2$. In case we specify to a dilatation $z'= \lambda z$, the prefactor is just $\lambda^{\Delta}$ hence $\Delta$ is called the \textit{scaling dimension}.\\

Conformal symmetry is very restrictive. It puts a lot of constraints on correlation functions. Therefore it is easier to ``solve" CFT's then the less constrained QFT's. The two-point function is even completely fixed by conformal symmetry. Suppose we are interested in $\langle O_{\Delta_1}(x_1,\bar{x_1})O_{\Delta_2}(x_2,\bar{x}_2)\rangle.$ Its behavior under translations fixes it to only depend on $\left| x_{1}-x_2\right|$. Dilatations then fix it to
\begin{equation}
 \langle O_{\Delta_1}(x_1,\bar{x_1})O_{\Delta_2}(x_2,\bar{x}_2)\rangle = C_{\Delta_1\Delta_2}\left| x_{1}-x_2\right|^{-\Delta_1-\Delta_2}.
 \label{eq:CFT2pt}
 \end{equation}
The behavior under special conformal transformations and a proper normalization then leads to $C_{\Delta_1\Delta_2}=\delta_{\Delta_1\Delta_2}$. The two-point function is completely fixed. Likewise the three-point function can be completely fixed up to a constant which is called the \textit{fusion coefficient} and which is theory dependent. The three point function reads
\begin{equation}
\langle O_{\Delta_1}(w_1)O_{\Delta_2}(w_2)O_{\Delta_3}(w_3)\rangle = \frac{C_{O_{\Delta_1}O_{\Delta_2}O_{\Delta_3}}}{\left| w_{12}\right|^{\Delta_1+\Delta_2-\Delta_3}\left| w_{13}\right|^{\Delta_1+\Delta_3-\Delta_2}\left| w_{23}\right|^{\Delta_2+\Delta_3-\Delta_1}},
\label{eq:CFT3pt}
\end{equation} 
The power of conformal field theories is that all higher point functions are fixed as soon as the fusion coefficients are determined. 
\subsection{AdS/CFT}
A bulk field in AdS $\phi(x,t,y)$ in Poincar\'e coordinates has the near boundary expansion~\cite{Aharony:1999ti}
\begin{equation}
 \phi(x,t,y\rightarrow 0) = \phi_0(x,t) y^{d-\Delta} + \tilde{\phi}(x,t) y^\Delta.
 \label{eq:scalexpand}
\end{equation}
The first term diverges near the boundary and when plugged into the scalar action, $\phi_0$ can be matched with the CFT source $J(x,t)$ dual to a scalar operator $O_\Delta (x,t)$. The second term vanishes near the boundary but contains the leading behavior of the scalar field in the absence of sources. Likewise it can be shown~\cite{Witten:1998qj,Balasubramanian:1998sn} that $\tilde{\phi} = \langle O_\Delta(x,t)\rangle$. These results are rederived in appendix~\ref{sec:app0}. The near boundary behavior further implies that in absence of a source 
\begin{equation}
G_{\partial\partial,\Delta} = \langle \phi(x_1,t_1,y_1\rightarrow 0)\phi(x_2,t_2,y_2\rightarrow 0)\rangle_{\text{AdS}} \sim\langle O_\Delta(x_1,t_1)O_\Delta (x_2,t_2)\rangle_{\text{CFT}}.
\end{equation}
where the $\sim$ symbol means that it matches the CFT two-point function up to a normalization factor, which is slightly tricky in AdS/CFT~\cite{Freedman:1998tz}.
The CFT two-point function matches a regularized version of the boundary-boundary propagator in AdS. An explicit check shows the result matches with the boundary limit of~(\ref{eq:bbprop}).

\section{HKLL}\label{sec:HKLL}
\subsection{A conceptual discussion of bulk reconstruction}
Because of the isomorphism between bulk and boundary operator algebras one expects any  local bulk operator $\phi(x,y)$ to be isomorphic to a combination of local CFT operators.  The HKLL method provides a way of expressing the local bulk operator through CFT operators in the large $N$ limit where $\phi$ satisfies the free equation of motion. We will implement the HKLL program in detail but conceptually the logic comes down to this: the scalar field solution can be expanded  as a linear combination of independent modes on AdS. In absence of sources, the boundary limit of the scalar is proportional to the vacuum expectation value of the dual CFT operator, so also the CFT operator can be expanded in terms of modes on AdS with the same coefficients as the bulk scalar field. Inverting the relation expresses the coefficients in terms of CFT local operators. Plugging this back into the mode expansion for the bulk scalar field results into an expression for the field in terms of boundary operators which can be schematically denoted as $\phi(x,t,y) = \int\limits  dx'dt' K(x,t,y; x',t')O_\Delta(x',t')$. $K$ is called the smearing function and will have support such that the operators $O_\Delta$ are spacelike separated from $\phi$. The smearing function formalism automatically allows the computation of bulk correlators from CFT ones.
\begin{align}
\langle \phi(x_1,y_1)\ldots \phi(x_n,y_n)\rangle_\text{AdS} = \int dx'_1 \ldots dx'_n  &K(x_1,y_1; x'_1)\ldots K(x_n,y_n; x'_n)\nonumber\\
\times
& \langle O(x'_1)\ldots O(x'_n)\rangle_\text{CFT}.
\end{align} 
\begin{figure}[h]
\centering
    \begin{subfigure}[h]{0.4\textwidth}            
 			\centering           
            \includegraphics[scale=0.25]{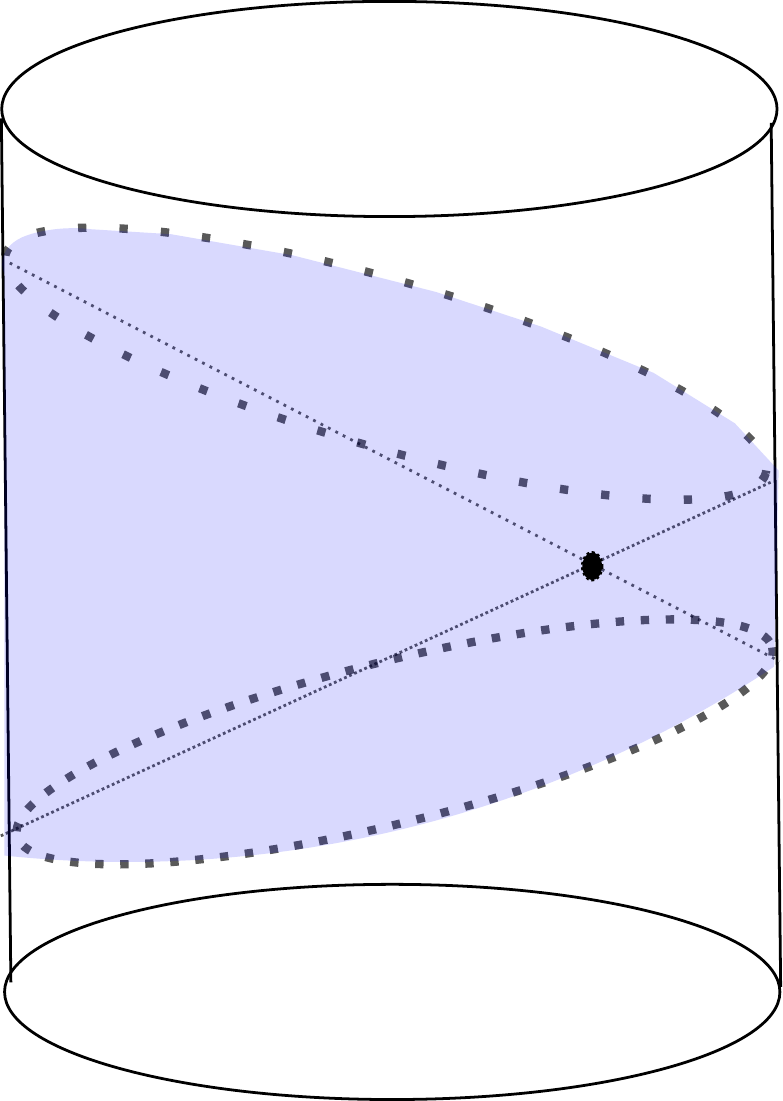}\quad
            \caption{}
            \label{fig:globalreconstruct}
    \end{subfigure}
	\begin{subfigure}[h]{0.4\textwidth}
			\centering
			\includegraphics[scale=0.25]{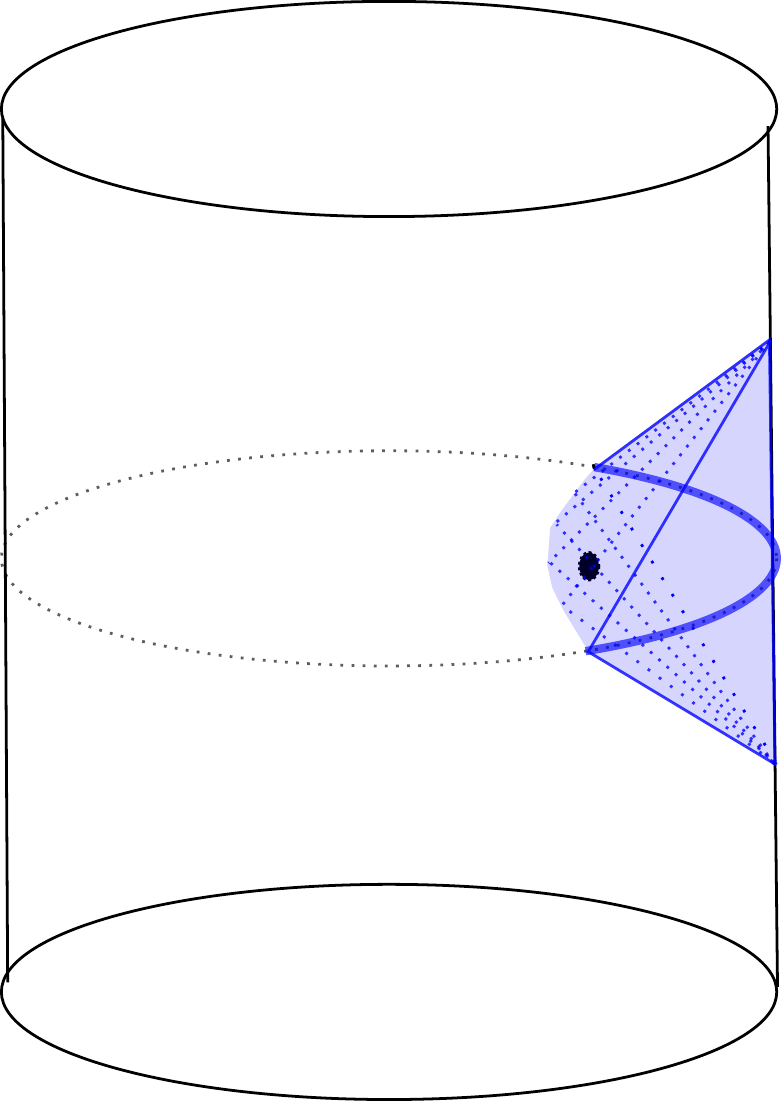}
			\caption{}
			\label{fig:rindlerreconstruct}
    \end{subfigure}%
	\caption{(a) sketches the region of support of the smearing function in global AdS coordinates, while (b) depicts the region of support of the smearing function on the rindler wedge of a particular boundary region $A$. The black dot indicates the position of the reconstructed bulk operator.}
\end{figure}

The set of boundary operators spacelike separated from the bulk operator, on which the smearing function has support, seems to depend on the choice of coordinate system. In global coordinates it naturally covers operators on the complete boundary spatial circle (see figure~\ref{fig:globalreconstruct}), while in Rindler coordinates it only covers the operators that are spacelike separated and contained in the Rindler patch. The Rindler patch is naturally associated to a boundary region that covers half of the spatial circle, but by an isometry the patch can be mapped to a patch that ends on any arbitrary spatial region $A$ on the boundary, as in figure~\ref{fig:rindlerreconstruct}.

\subsection{Construction of the smearing function in global $AdS_{d+1}$}
Here we will construct the smearing function in global coordinates. We will first work for any $d$ and then specialize to $d=2$. We start from the metric in global coordinates
\begin{equation}
ds^2 = \frac{R^2}{\cos^2\rho} \LF - d\tau^2  + d\rho^2 + \sin^2\rho d\Omega_{d-1}^2\RF
\end{equation}
or specializing to $AdS_3$ it reduces to~(\ref{eq:globalcoord}).
The free scalar field EOM in global coordinates is 
\begin{equation}
-\partial_\tau^2\phi + \partial_\rho^2\phi + \frac{1}{\sin\rho\cos\rho}\partial_\rho \phi + \frac{\hat{L}^2}{\sin^2\rho} \phi - \frac{\Delta(\Delta-d)}{\cos^2\rho}\phi = 0.
\end{equation} 
Solving the EOM by expanding into eigenmodes renders
\begin{equation}
\phi(\tau,\rho,\Omega) = \sum\limits_{n=0}^{+\infty} \sum\limits_{l,m} a_{nl} \Ym_l(\Omega) e^{-i (2n+l+\Delta)\tau} \cos^\Delta\rho \sin^l \rho P_n^{(\Delta-\frac{d}{2},l+\frac{d}{2}-1)}(-\cos 2\rho) + \text{c.c.}
\end{equation}
with $\Ym_l$ the $l$'th spherical harmonic and $P_n^{(\alpha,\beta)}$ the Jacobi polynomial. In the case of $d=2$ this becomes
\begin{equation}
\phi(\tau,\rho,\Omega) = \sum\limits_{n=0}^{+\infty} \sum\limits_{l,m} a_{nl} e^{il\varphi} e^{-i (2n+l+\Delta)\tau} \cos^\Delta\rho \sin^l \rho P_n^{(\Delta-1,l)}(-\cos 2\rho) + \text{c.c.}
\end{equation}
In the absence of sources, the boundary value of the field is
\begin{equation}
\langle O_\Delta (\tau,\Omega) \rangle = \tilde{\phi}(\tau,\Omega) = \lim\limits_{\rho\rightarrow \frac{\pi}{2}} \frac{\phi (\tau,\rho,\Omega)}{\cos^\Delta \rho}.
\end{equation}
Notice that instead of working with operators, we are working with classical fields. we could lift the boundary condition to an operator equation by dropping expectation values and restoring the hats on $O_\Delta$ and $\phi$.\footnote{In doing so, the operator $\phi$ will actually be state dependent.}\\

Let us first reconstruct the operator at the center of AdS. Because the field solution in the origin should be spherically symmetric only the $l=0$ mode can contribute and the solution reduces to
\begin{equation} 
\phi(\tau,\rho=0,\Omega) = \sum\limits_{n=0}^{+\infty} a_n e^{-i(2n+\Delta)\tau} P_n^{(\Delta-\frac{d}{2},\frac{d}{2}-1)} (-1) + \text{c.c}.
\label{eq:origin}
\end{equation}
The $l=0$ mode part of the boundary field on the other hand equals
\begin{align}
\tilde{\phi}^s (\tau,\Omega) &= \sum\limits_{n=0}^{+\infty} a_n e^{-i (2n+\Delta)\tau} P_n^{(\Delta -\frac{d}{2},\frac{d}{2}-1)}(+1) +\text{c.c.},\\
&= \tilde{\phi}_+^s (\tau) + \tilde{\phi}_-^s (\tau)
\end{align}
when explicitly splitting up the field into positive and negative frequency modes. This relation can be solved for the coefficients $a_n$:
\begin{equation}
a_n = \frac{1}{\pi \text{Vol}(S^{d-1}) P_n^{(\Delta- \frac{d}{2}, \frac{d}{2}-1)} (+1)} \int\limits_{-\frac{\pi}{2}}^{+\frac{\pi}{2}} d\tau \int d\Omega \sqrt{g_\Omega} e^{i(2n+\Delta)\tau} \tilde{\phi}_+^s (\tau,\Omega)
\end{equation}
and plugging back into~(\ref{eq:origin}) results in
\begin{align}
\phi(\tau',\rho=0,\Omega') &= \int\limits_{-\frac{\pi}{2}}^{\frac{\pi}{2}} d\tau \int\limits d\Omega K_+ (\tau,\Omega; \tau',\rho=0,\Omega') \tilde{\phi}_+^s (\tau,\Omega) + \text{c.c},\label{eq:ReconstructCenter}\\
\text{with} \quad K_+ &= \frac{1}{\pi \text{Vol}(S^{d-1})}\sum\limits_{n=0}^{+\infty} e^{i (2n+\Delta)\tau} \frac{ P_n^{(\Delta-\frac{d}{2},\frac{d}{2}-1)}(-1)}{P_n^{(\Delta-\frac{d}{2},\frac{d}{2}-1)}(+1)}.
\end{align}
The Jacobi polynomials reduce to simple Gamma functions
\begin{align}
P_n^{(\Delta-\frac{d}{2}, \frac{d}{2}-1)} (-1) &= \frac{(-1)^n (d+1)_n}{n!},\\
P_n^{(\Delta-\frac{d}{2}, \frac{d}{2}-1)} (+1) &= \frac{ \LF \Delta - \frac{d}{2}+1\RF_n}{n!},
\end{align}
with $(x)_n= \Gamma(x+n)/\Gamma (x)$ the Pochhammer symbol. The smearing function becomes a sum over $n$ with Gamma function coefficients forming the standard hypergeometric series, and can be easily resummed to
\begin{equation}
K_+ = \frac{ e^{i\Delta \tau}}{\pi \text{Vol}(S^{d-1})} \left._2 F \right._1\LF 1,\frac{d}{2}; \Delta-\frac{d}{2}+1,-e^{2i\tau}\RF.
\end{equation} 
For even dimensional AdS (i.e. odd $d$) and with the aid of hypergeometric identities, the smearing function becomes
\begin{align}
K_+  =& \frac{e^{i\Delta \tau}}{\pi \text{Vol}(S^{d-1}) }\left\{ \frac{\LF \Delta -\frac{d}{2}\RF\Gamma \LF \frac{d}{2}-1\RF}{\Gamma\LF \frac{d}{2}\RF} e^{-2i\tau} \left._2 F\right._1\LF 1,1+\frac{d}{2}-\Delta, 2-\frac{\Delta}{2},-e^{-2i\tau}\RF\right.\nonumber\\
& \left. + \frac{\Gamma\LF \Delta-\frac{d}{2}+1\RF \Gamma \LF 1- \frac{d}{2}\RF}{\Gamma\LF \Delta - d+1\RF} e^{-id\tau}\left._2 F \right._1 \LF d-\Delta,\frac{d}{2},\frac{d}{2},-e^{-2i\tau}\RF\right\}.
\end{align} 
The above identities cannot be used for even $d$ because of singularities that apppear in the Gamma functions. $K_+$ is the smearing function by which an operator at the center is reconstructed using positive frequency modes. The negative frequency modes then automatically appear in the complex conjugated part. Reality of the bulk field then requires $K_+^*=K_-$ with $K_-$ the smearing function when we would have used negative frequency modes to express the coefficients $a^*_n$. Notice that the range of integration in~(\ref{eq:ReconstructCenter}) is over all boundary operators that are spacelike separated from the bulk operator at the center.\\

The smearing function is actually more of a distribution, it is only defined when integrated against the boundary operators. Said otherwise, $K_+$ is not unique and can be shifted by terms which vanish when integrated against the boundary operators and
\begin{equation}
K_+ \rightarrow K_+ + e^{i\Delta \tau} \sum\limits_{n=1}^{+\infty} c_n e^{2in\tau},
\end{equation}
for any coefficient $c_n$. Using this gauge freedom to fix $K_+$ the first term can be canceled and one is left with
\begin{equation}
K_+ = \frac{\Gamma\LF \Delta-\frac{d}{2}+1\RF \Gamma \LF 1- \frac{d}{2}\RF}{\pi \text{Vol}(S^{d-1})\Gamma\LF \Delta - d+1\RF} \LF 2\cos\tau\RF^{\Delta-d}.
\end{equation}
This is a real expression so since $K_-= K_+^*$ we have that $K=K_+=K_-$. For a bulk point at the origin $\cos\tau = \sigma\cos\rho$ with $\sigma$ the invariant distance function between the bulk and boundary point. This allows a simple extension to determine the smearing function for any bulk point. Anti- de Sitter space can be viewed as the goup manifold of the AdS isometry group $SO(d,2)$. Applying an isometry will generically map the bulk point at the origin to another point of AdS. Because of the isometry covariance of AdS, the smearing function can only be a function of the invariant distance\footnote{In very much the same way as the AdS propagator can only be a function of the invariant distance.} between the respective bulk and boundary point. We arrive at
\begin{equation}
K = \frac{\Gamma\LF \Delta-\frac{d}{2}+1\RF \Gamma \LF 1- \frac{d}{2}\RF}{\pi \text{Vol}(S^{d-1})\Gamma\LF \Delta - d+1\RF} \LF 2\sigma\cos\rho\RF^{\Delta-d}
\end{equation}
for any bulk point. A similar procedure can be followed for even $d$. The resulting smearing function in that case is
\begin{equation}
K= \frac{ (-1)^{\frac{d-2}{2}} \Gamma\LF \Delta-\frac{d}{2}+1\RF}{ 2\pi^{1+\frac{d}{2}} \Gamma \LF \Delta-d +1\RF} \lim\limits_{\rho\rightarrow \frac{\pi}{2}} \LF 2\sigma \cos\rho\RF^{\Delta-d}\log\LF \sigma\cos\rho\RF.
\end{equation}
The smearing function has support on the  boundary points that are spacelike separated from the bulk insertion point. It cannot have support on timelike separated points because the CFT operators on timelike separated points generically do not commute with the bulk operator. This is basically stating causality.

\subsection{Recovering local bulk correlators in Poincar\'e coordinates}
Given an HKLL construction of local bulk operators it should be possible to compute bulk correlation functions from the corresponding CFT correlation functions. Because the analytic tractability depends a lot on the specific set of parameters like $\Delta$ and $d$ and on the coordinate system, we will study bulk correlators only in the specific case of $d=2$, $\Delta=2$ (which means that we have a massless bulk scalar) in Poincar\'e coordinates. The particular bulk correlator that we will compute is the correlator of a bulk with a boundary field, which should reproduce the bulk-boundary propagator in AdS. As was derived in~\cite{Hamilton:2006fh,Hamilton:2006az} and we will rederive in appendix~\ref{sec:app1}, the bulk field as a smeared boundary operator can be written in Poincar\'e coordinates on a patch that effectively becomes of zero size as one approaches the boundary.
\begin{equation}
\phi(t,x,y) = \frac{\Delta-1}{\pi} \int\limits_{t^{'2}+y'^{2}\leq y^2} dt' dy' \LF \frac{y^2 -t^{'2} - y^{'2}}{y}\RF^{\Delta-2} \tilde{\phi}^P (t+t',x+iy').
\end{equation}
Here the label $P$ stands for Poincar\'e. Notice that the modes are smeared in the imaginary direction and we have implicitly analytically continued the boundary fields. From this we compute the bulk-boundary $2$ point correlator:
\begin{equation}
\langle \phi(t,x,y) \tilde{\phi}^P(0,0)\rangle = \frac{\Delta-1}{\pi} \int\limits_{t^{'2}+y'^{2}\leq y^2} dt' dy' \LF \frac{y^2 -t^{'2} - y^{'2}}{y}\RF^{\Delta-2} \langle\tilde{\phi}^P (t+t',x+iy')\tilde{\phi}^P(0,0)\rangle
\end{equation}
where we used translation invariance on the boundary to put one operator at the origin $(t,x)=(0,0)$. The boundary CFT two-point function~(\ref{eq:CFT2pt}) in Poincar\'e coordinates is
\begin{equation}
\langle\tilde{\phi}^P (t+t',x+iy')\tilde{\phi}^P(0,0)\rangle =  \frac{1}{2\pi} \frac{ 1}{ \LF (t+t')^2 - (x+iy')^2\RF^\Delta}.
\end{equation}
A map to polar coordinates transforms the integral into
\begin{equation}
\langle \phi(t,x,y) \tilde{\phi}^P(0,0)\rangle = \frac{\Delta -1}{\pi^2} \int\limits_{0}^y dr \LF \frac{y^2-r^2}{y}\RF^{\Delta- 2} \oint\limits_{\left|z\right|=1} \frac{ z dz}{ 2\pi i \LF t+x+ry\RF^\Delta \LF y(t-x)+r\RF^\Delta}.
\end{equation}
Generically one will have to deal with the branch cut structure of the contour integral, but for the special case of a massless scalar field in $d=2,\Delta=2$ the branch cuts become ordinary poles. 
We distinguish two regimes.
\begin{itemize}
\item In the case of $\left|t+x\right|\geq y$ and $\left|t-x\right| \geq y$, one always encircles the pole at $z= r/(x-t)$ such that the residue theorem implies:
\begin{equation}
\langle \phi(t,x,y) \tilde{\phi}^P(0,0)\rangle = \frac{1}{2\pi} \frac{ y^2}{\LF t^2-x^2-y^2\RF^2}.
\end{equation}
This is the bulk-boundary propagator as expected.
\item In the other case of $\left|t+x\right|\leq y$ or $\left|t-x\right|\leq y$ one of the poles crosses the contour, so in order to have a well defined integral, the contour needs to be deformed in such a way that the pole at $z=r/(x-t)$ is always being encircled. Then the bulk-boundary propagator will always be reproduced.
\end{itemize}
\subsection{Non-vanishing commutators}
A truly local bulk-operator has to commute with spacelike separated operators. To illustrate that this does not happen in an HKLL reconstruction, we compute the following correlation function
\begin{align}
&\langle \LT \phi(x_0,t_0,y),O(x_1,t_1)\RT O(x_2,t_2)\rangle \nonumber\\
&= \langle \phi(x_0,t_0,y)O(x_1,t_1) O(x_2,t_2)\rangle - \langle  O(x_1,t_1) \phi(x_0,t_0,y) O(x_2,t_2)\rangle.
\end{align}
where the bulk field is massless and the operator $O$ is the boundary operator associated with that field. $t$ denotes the Lorentzian time, so the ordering of operators is important. The correlator can be computed by explicitly performing the expansion as a convolution with the smearing function and using the CFT three-point function~(\ref{eq:CFT3pt}) the correlation function becomes proportional to
\begin{align}
 &\langle \phi(x,t,y)O(x_1,t_1) O(x_2,t_2)\rangle\nonumber\\
 &\sim \int\limits_{y^{'2} + t^{'2}\leq y^2} dt' dy' \frac{ 1}{ \LT (t_1-t_0-t')^2-(x_1-x_0-iy')^2\RT \LT (t_2-t_0-t')^2 - (x_2-x_0 - iy')^2\RT.} 
\end{align}
We have neglected the fusion coefficients here which are actually $O(1/N)$.\footnote{Three point functions can be studied holographically by a Witten diagram~\cite{Witten:1998qj} of a vertex with three legs in the bulk. Such a Witten diagram is only non-zero if there are three point interactions in the scalar Lagrangian. So what we have actually done in assuming a free scalar, is neglecting all higher order interactions in the scalar Lagrangian by assuming they are subleading compared to the kinetic and mass term.}  Going to coordinate $(t',y')=(r\cos\theta,r\sin\theta)$ and then defining $z=e^{i\theta}$, the integral can be turned into
\begin{align}
\int\limits_0^y rdr \oint\limits_{\left|z\right|=1} \frac{zdz}{ \LT x_{10}^+ -rz)(zx_{10}^- -r)(x_{20}^+ -zr)(zx_{20}^- - r)\RT}
\end{align}
with $ x_{kl}^{\pm} = (t\pm x)_k - (t\pm x)_l$. The contour encircles poles at $z=\frac{r}{x_{10}^-}$ and $z=\frac{r}{x_{20}^-}$ so by the residue theorem we eventually arrive at
\begin{align}
 &\langle \phi(x,t,y)O(x_1,t_1) O(x_2,t_2)\rangle\nonumber\\
 &\sim \frac{ 1}{\LT (x_1-x_2)^2 - (t_1-t_2)^2\RT} \ln\LT\frac{ (y^2-x_{10}^+x_{10}^-) (y^2-x_{20}^+x_{20}^-)}{ (y^2-x_{20}^+x_{10}^-) (y^2-x_{10}^+x_{20}^-)}\RT.
\end{align}
The point is that the correlator contains unwanted singularities at $y^2 = x_{20}^{\pm}x_{10}^\mp$ next to the usual Lorentzian lightcone singularities. These singularities imply that an $i\epsilon$ prescription is needed which is different for different orderings of the operators inside the correlator. Therefore the commutator doesn't vanish. To resolve this issue we notice that we have assumed that the bulk field $\phi$ is a convolution of boundary operators which are primary operators in such a way that $\phi$ satisfies the bulk equation of motion. In order not to modify the leading near-boundary behavior of $\phi$ only operators of higher conformal weight, appropriately smeared, can still be added to its definition. From demanding that the set of added operators should remove the log-type singularities of the correlator it follows that the only possible terms that can be added are the multitrace operators like $O_{\Delta_k}\equiv  : O_\Delta \overset{\leftrightarrow}{\partial}^k O_\Delta:$ with $\Delta_k = 2\Delta +2k$. Multitrace operators correspond to multiparticle states in the bulk. We define the bulk field as
\begin{equation}
\phi = \int\limits K_\Delta O_\Delta + \sum\limits_k d_k \int K_{\Delta_k} O_{\Delta_k}.
\end{equation}
The double trace operators actually make sure that the bulk field is part of a consistent interacting field theory, which is not a free field theory but a generalized free field theory~\cite{Balasubramanian:1999ri,ElShowk:2011ag,Kabat:2013wga}. Indeed, the double trace operators give rise to a potential term in the equations of motion
\begin{align}
\LF \Box - m_\Delta^2\RF \phi &= \int (\Box-m_\Delta^2)K_\Delta O_\Delta + \sum\limits_{k} d_k (\Box - m_\Delta^2) K_{\Delta_k} O_{\Delta_k},\\
&= \sum\limits_k d_k (m_{\Delta_k}-m_\Delta^2)\int K_{\Delta_k}O_{\Delta_k},
\end{align}
by using that the smearing function satisfies the homogeneous wave equation. On the right hand side an effective potential has been generated by the double trace terms. These effects are $1/N$ compared to the leading free equations of motion.
\subsection{The problems with HKLL}
The HKLL method seems to give a definition of bulk operators intrinsic to the boundary CFT, but in order to know the smearing function one needs to solve the bulk equations of motion. This presumes an existing bulk geometry in which the scalar field acts as a probe. Tied to this we notice that the HKLL method actually links bulk fields to vacuum expectation values of CFT operators, which after promoting the equality to an operator equality, the bulk operator becomes state dependent.  CFT operators on the other hand are well-defined elements of the operator algebra and because of the isomorphism between bulk and boundary operators, so should the bulk operators. We would therefore like to remove the state dependence and have well defined background independent bulk operators. On top of that, note that the bulk field is classical, which is only valid at leading order in $N$. At finite $N$ quantum corrections modify the operator, leading to an interacting scalar field. HKLL has initiated a program to define bulk operators perturbatively in $N$ in such a way that bulk correlators are well defined, but ideally they should be defined non-perturbatively.\\

  A local bulk scalar field in global coordinates still depends on all CFT operators at a fixed time, apparently even when the bulk field is pushed to the boundary, which is at odds with the local boundary condition that we imposed on the field. Rindler coordinates solve this problem but at the same time pose another paradox. How can both methods of reconstruction be reconciled and describe the same bulk operator? This has led Almheiri, Dong and Harlow, to propose a link between AdS/CFT and Quantum Error Correction (QEC), the underlying idea being that bulk operators are not unique but depend on the wedge of reconstruction~\cite{Almheiri:2014lwa} which will be the subject of section~\ref{sec:QEC}.\\

The constructed operators with HKLL explicitly depend on the coordinate system and are not diffeomorphism invariant. The diffeomorphism symmetry is a gauge symmetry of the bulk and every element of the operator algebra should be diffeomorphism invariant. The operators could be made invariant by dressing them gravitationally, as we will explain in more detail in section~\ref{sec:gauginv}. The dressing is a subleading process in $N$ and does not influence the leading order construction of HKLL.\\

Concludingly, the HKLL method provides a nice way of reconstructing bulk operators from CFT operators but has a couple of conceptual problems associated with it. The next sections will try to zoom in on these problems and discuss possible resolutions. Unfortunately, anno 2017, there's no resolution yet that solves all of the above mentioned problems for generic holographic theories.

\section{Quantum Error Correction}
\label{sec:QEC}
\begin{figure}[h]
\centering
\includegraphics[scale=0.4]{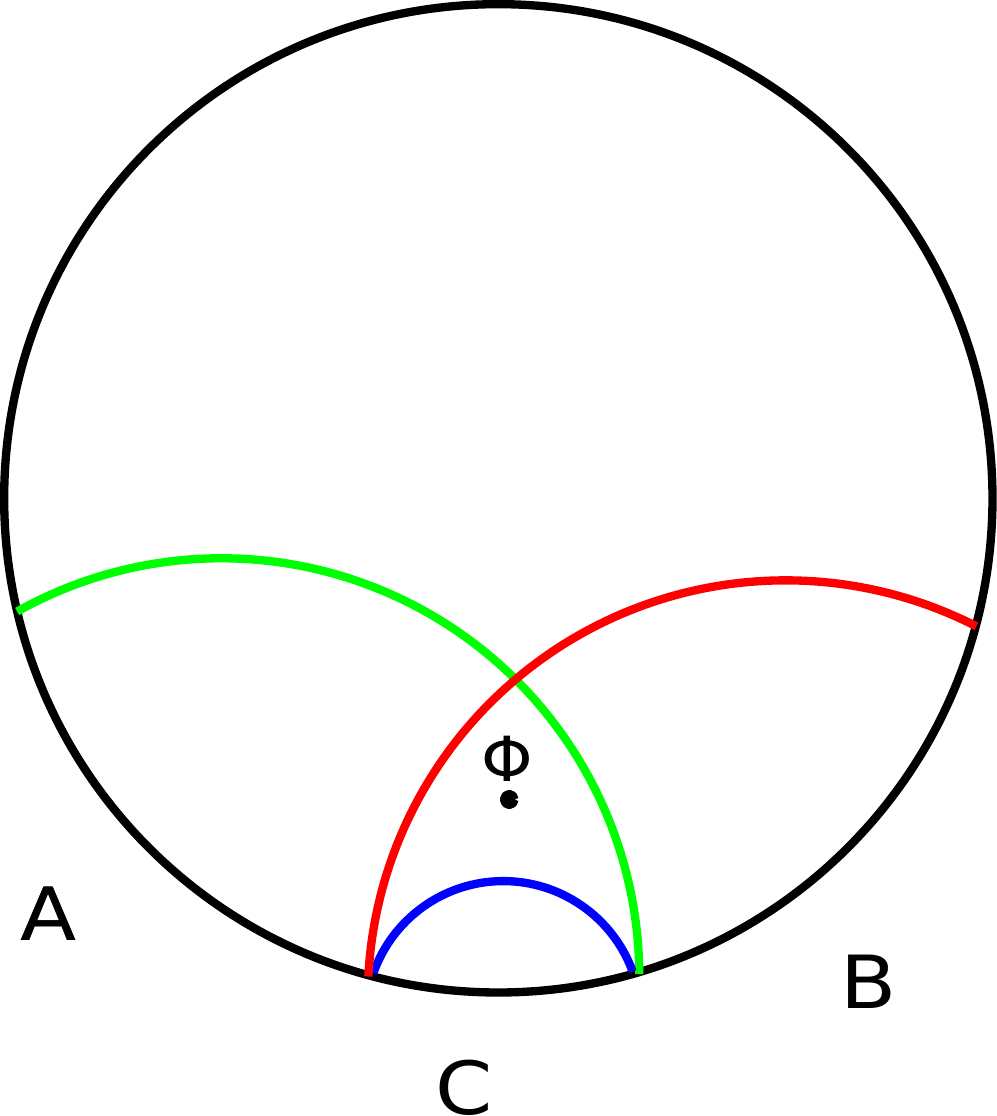}
\caption{Reconstruction on overlapping boundary regions $A$ and $B$. The minimal surfaces anchored on A, B and C are respectively the green, red and blue curves.}
\label{fig:paradox}
\end{figure}

Let us specialize here to reconstruction on $\text{AdS}_3$. We want a reconstruction such that the smearing is more and more local as the bulk operator approaches the boundary. As we have seen this is not the case in global coordinates but HKLL already noted that it can be done in Rindler coordinates, by smearing over the Rindler patch~\cite{Hamilton:2006fh,Hamilton:2006az}. However, the construction in the Rindler patch is still coordinate dependent. To resolve this we define the \textit{causal} wedge. Take a CFT Cauchy surface $\Sigma$ with $A\subseteq\Sigma$. We define the boundary domain of dependence $D[A]$ as the region in the boundary such that every inextendible causal curve\footnote{i.e. curve with timelike or lightlike tangent vectors.} that passes through $D[A]$ also intersects $A$. We define the causal future/past of $D[A]$ as $\mathcal{J}^{+}\LF D[A]\RF / \mathcal{J}^{-}\LF D[A]\RF$. The causal wedge is their intersection $\mathcal{W}[A] = \mathcal{J}^{+}\LF D[A]\RF \cap \mathcal{J}^{-}\LF D[A]\RF$. When $\Sigma$ is the $t=0$ slice (with $t$ the boundary time) and $A$ is a semicircle, then $\mathcal{W}[A]$ is the Rindler patch of $AdS_3$. On the Rindler patch there exists an HKLL smearing function in Rindler coordinates. Next, one can use isometries of AdS to map this wedge to any causal wedge $\mathcal{W}[A]$, as in figure~\ref{fig:rindlerreconstruct}. This proves that an HKLL procedure exists for reconstruction in any causal wedge\footnote{The literature contains discussions on the possibility of reconstruction in a bigger wedge called the entanglement wedge. We will however not go into the details of this construction as for vacuum AdS the causal and entanglement wedge coincide~\cite{Hubeny:2012wa,Headrick:2014cta}.}. If the boundary subregion $A$ is such that $\phi \in \mathcal{W}[A]$ then a reconstruction in terms of $O\in \Am(A)$ exists. We have not introduced coordinates along the way, so reconstructability is coordinate independent. However a new puzzle appears. In figure~\ref{fig:paradox} the timeslice of $AdS$ is drawn. Take an operator $\phi$ such that it is contained in the causal wedges of both $A$ and $B$ but not in their intersection. Then we would say that $\phi$ can be reconstructed from both $A$ and $B$ but not from their intersection $C\equiv A\cap B$. However, since operators on $A$ and $B$ are spacelike separated, they must commute, and therefore the only possibility is for $\phi$ to have non-trivial support on the boundary when the operators are part of $A\cap B$. This is a paradox!\\
 
The hidden assumption here is that the operator is independent of the chosen subregion, so the only way out is to conclude that the bulk operator depends on the choice of subregion. As we will discuss next, this is precisely a signature that the bulk behaves as a quantum error correcting code.\\

\subsection{$3$-qutrit model}
Take a spin model consisting of $3$ qutrits and think of the qutrits as making up the physical degrees of freedom on the boundary. So in this case the would-be boundary consists of only $3$ points.\\ 

Say Alice wants to send a qutrit to Bob, but along the way it can be lost or erased. How can Alice reliably send information to Bob? Say she wants to send the state
\begin{equation}
\ket{\psi} = \sum\limits_{i=0}^2 a_i \ket{i}.
\end{equation}
What she can do is to send $3$ qutrits instead:
\begin{equation}
\left\{ \begin{matrix}
\ket{0}\rightarrow \ket{000},\\
\ket{1}\rightarrow \ket{111},\\
\ket{2}\rightarrow \ket{222}
\end{matrix}
\right.
\end{equation}
and send the following state
\begin{align}
\ket{\tilde{\psi}} &= \sum\limits_{i=0}^{2} a_i \ket{\tilde{i}},\\
\text{with} \quad & \left\{
\begin{matrix}
\ket{\tilde{0}} = \frac{1}{\sqrt{3}} \LF \ket{000} + \ket{111} + \ket{222}\RF,\\
\ket{\tilde{1}} = \frac{1}{\sqrt{3}}\LF \ket{012}+ \ket{120} + \ket{201}\RF,\\
\ket{\tilde{2}} = \frac{1}{\sqrt{3}}\LF \ket{021} + \ket{102} + \ket{210}\RF.
\end{matrix}
\right.
\end{align}
The advantage of copying the basisstate and sending the new state is that Bob can still reliably reconstruct the state even if a qubit has been lost. We assume the probability of losing a qubit is of $O(\epsilon)$ with $\epsilon \ll 1$ such that the probability of losing two qubits is $O(\epsilon^2)$ and negligible. The basisstates $\ket{i}^{\otimes 3}$ form the \textit{physical} Hilbert space, while the states $\ket{\tilde{i}}$ form the \textit{logical} Hilbert space or code subspace.  Say that the third qutrit can be lost and Bob only has acces to the first two qutrits. Bob can always apply a unitary transformation on the first two qutrits from which he can distill what the physical state is. He can apply the unitary
\begin{equation}
\LF U_{12}\otimes I_3 \RF \ket{\tilde{i}} = \ket{i} \otimes \frac{1}{\sqrt{3}}\LF \ket{00} + \ket{11} + \ket{22}\RF, 
\end{equation}
and therefore
\begin{equation}
\LF U_{12}\otimes I_3 \RF \ket{\tilde{\psi}} = \ket{\psi} \otimes \frac{1}{\sqrt{3}}\LF \ket{00} + \ket{11}+\ket{22}\RF.
\end{equation}  
By applying the unitary he can read out $\ket{\psi}$. Bob is able to distill the physical information because the logical qutrits $\ket{\tilde{i}}$ are maximally entangled such that the information was stored non-locally through the entanglement. Even if a qutrit is erased, the information is not lost. Clearly although the state that was sent contained $3$ physical qutrits, only one logical unit of information was sent.\\

Moreover an isomorphism between operators on the physical Hilbert space and the code subspace exists. Say that $O$ is an operator acting on a single qutrit with action 
\begin{equation}
O\ket{i} = \sum\limits_j \LF O\RF_{ji}\ket{j}
\end{equation}
then an operator can be constructed that has the same action on the code subspace:
\begin{equation}
\tilde{O}\ket{\tilde{i}} \equiv \sum\limits_j  \LF O\RF_{ji}\ket{\tilde{j}}.
\end{equation}
One can even construct an operator that has the same action on the code subspace as $O$ on the physical one, but only has non-trivial support on the first two qutrits.
Say $O_{12} = U_{12}^\dagger O U_{12}$. This operator has the action on the code subspace
\begin{align}
O_{12}\ket{\tilde{i}} &= U_{12}^\dagger O U_{12} \ket{\tilde{i}},\nonumber\\
&= U_{12}^\dagger O  \ket{i} \otimes \frac{1}{\sqrt{3}}\LF \ket{00} + \ket{11} + \ket{22}\RF,\nonumber\\
&= \sum\limits_j \LF O \RF_{ji} U_{12}^\dagger \ket{j} \otimes \frac{1}{\sqrt{3}}\LF \ket{00} + \ket{11} + \ket{22}\RF,\nonumber\\
&= \sum\limits_{j} \LF O\RF_{ji}\ket{\tilde{j}}.
\end{align}
The operator $O_{12}$ only acts on the first two qutrits but has an action that is equivalent to the action of $O$ on the physical subspace. We could have done a similar procedure constructing $O_{13}$ and $O_{23}$ instead. The conclusion is that we have found $3$ different operators that have support on different sets of qutrits but nevertheless have the same action on the code subspace.
\subsection{AdS/CFT as quantum error correction}
Think of the boundary Hilbert space as the physical Hilbert space\footnote{In the $3$-qutrit example, the space spannend by $3$ physical qutrits.}. The paradox of reconstruction of an operator $\phi$ that is in both $A$ and $B$ but not in the intersection could be resolved by thinking of the bulk Hilbert space as the code subspace. What we then mean by an operator $\phi(P)$ at a bulk point $P$ is actually either $\phi_{12}(P), \phi_{23}(P),\phi_{13}(P)$ depending on the wedge of reconstruction. Note that the bulk cannot distinguish between these three operators and they all have the same action on the bulk, which is reminiscent of the quantum error correcting properties of the bulk. This resolves the paradox. We cannot reconstruct $\phi$ from the intersection of $A$ and $B$ because $\phi_{A}$ and $\phi_B$ are genuinely different operators, even though they have the same action on the bulk Hilbert space. Stating that the wedge of reconstruction has to be taken larger and larger as the bulk operator moves inward radially, can be rephrased in quantum information language as saying that logical bulk operators represent logical operations which become better and better protected as the operator is moved inwards radially into the bulk~\cite{Almheiri:2014lwa}.\\

We have thought of the bulk Hilbert space here as the space of states of a field on a fixed background. In reality, the bulk Hilbert space contains many states which completely deform the background e.g. black hole states. The code subspace therefore does not agree with the full bulk Hilbert space but rather with an effective space of states where backreaction is neglected:
\begin{equation}
\ket{\Omega}, \phi_i(P_1)\ket{\Omega}, \phi_{i}(P_1)\phi_{j}(P_2)\ket{\Omega},\ldots
\end{equation}
where we have to cut off the number of field insertions at some large enough $n$ because \newline $\phi_{i_1}(P_1)\ldots\phi_{i_n}(P_n)\ket{\Omega}$ would eventually acquire a large enough energy to start deforming the background and our effective Hilbert space picture would break down. The problem of background dependence of bulk reconstruction has therefore not been solved by the QEC model.
\subsection{Tensor networks for AdS/CFT}
The quantum error correcting properties of AdS/CFT could be much better explained if we have a toy model at our disposal which characterizes the microscopics of the state. This is what has been aimed for in the literature~\cite{Pastawski:2015qua,Yang:2015uoa,Hayden:2016cfa} on holographic tensor networks and holographic codes. Entanglement between the microscopic degrees of freedom plays a crucial role in the encoding of the network and in the reconstruction of the geometry. Holographic codes therefore provide a toy model for studying the microscopic representation of entanglement of the CFT through a bulk geometry.\\

\begin{figure}[h]
\centering
\includegraphics[scale=0.3]{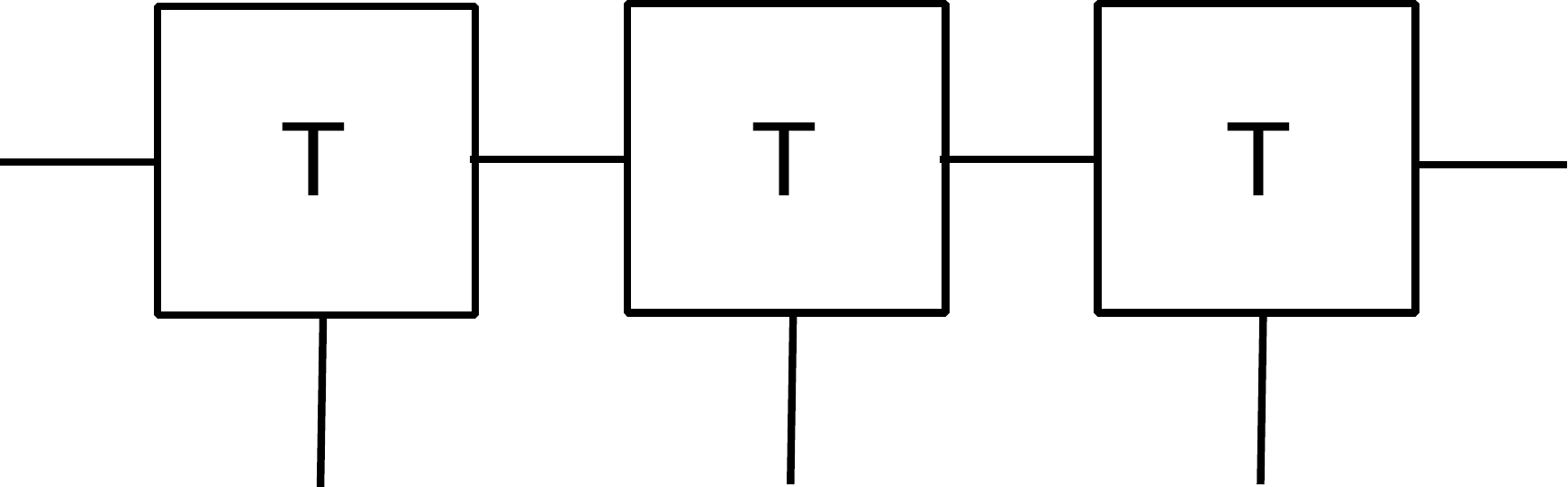}
\caption{A $d=1$ tensor network of 3 blocks. The contracted legs represent contracted indices. The free legs represent uncontracted legs, which are contracted in the end with basisstates in order to produce the full state.}
\label{fig:mps}
\end{figure}

Before specifying to holographic tensor networks, let's say just a few words about tensor networks in general. Consider a discretization of a $d=1+1$ boundary. Consider a timeslice and put spins on each point of the $d=1$ spatial lattice. A general state on the circular lattice can be decomposed into a basis as
\begin{equation}
\ket{\psi} = \sum\limits_{a_1,\ldots,a_n} T_{a_1\ldots a_n}\ket{a_1,\ldots,a_n}
\end{equation}
where each $a_i=1,\ldots,\chi$ with $\chi$ the dimension of the spin. The basis decomposition of a state defines an $n$-dimensional tensor with $n$ input legs. To reduce the dimension of this tensor we can make the ansatz that it is built from local tensors with $3$ indices as in figure~\ref{fig:mps} such that
\begin{equation}
\ket{\psi}\sum\limits_{a_1,\ldots,a_n, c_1\ldots,c_n} T_{a_1}^{c_n,c_1}T_{a_2}^{c_1,c_2}\ldots T_{a_n}^{c_{n-1},c_n} \ket{a_1,\ldots,a_n}.
\end{equation}

These chains form the so called \textit{matrix product states} and it turns out that this ansatz allows for an efficient computation of the groundstates of $1d$ gapped local Hamiltonians~\cite{Ostlund:1995zz,Vidal:2003lvx,Cirac:2004}.\\

Computing ground states of gappless systems can be done using the \textit{Multiscale Entanglement Renormalization Ansatz} (MERA)~\cite{Vidal:2007hda}. The idea is very similar. One starts with a state that is a tensor contraction of basisstates in terms of local degrees of freedom where also the tensors are local and have a bond dimension. This way of contracting allows nearest neighbor spins to be entangled. Then using a renormalization procedure consisting of isometry and disentangling operators, the ansatz state is effectively coarse grained to the ground state of the Hamiltonian. Information about the Hamiltonian is encoded in the isometric and disentangling operators. It turns out that the RG direction of this network has similar properties as the radial direction in holography~\cite{Swingle:2009bg,Swingle:2012wq} (which has been associated with renormalization flow already from the early days of AdS/CFT~\cite{Maldacena:1997re}). This further motivates building a tensor network structure that mimics the AdS geometry.\\

A first real attempt to build a tensor network that has the same hyperbolic structure as AdS and posesses quantum error correcting properties is the so-called HaPPY network\footnote{After Harlow, Pastawski, Preskill and Yoshida~\cite{Pastawski:2015qua}.}.
\begin{figure}[h]
\centering
\begin{subfigure}[h]{0.4\textwidth}
	\centering
	\includegraphics[scale=0.2]{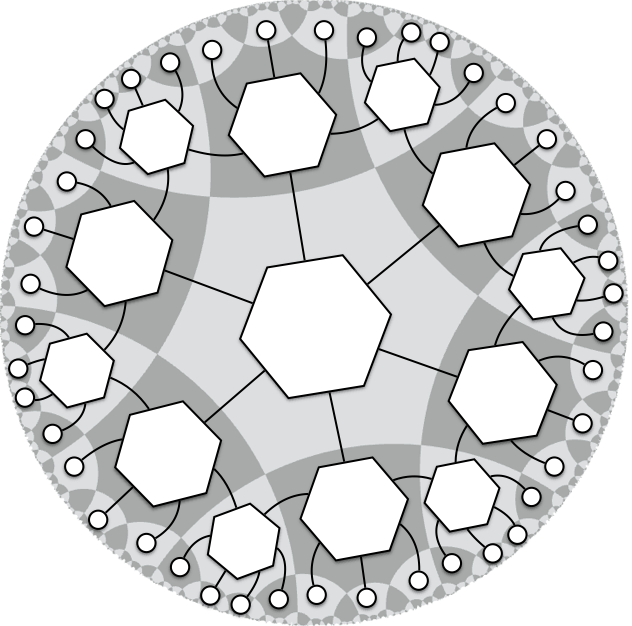}\quad
	\caption{}
	\label{fig:hexagon1}
\end{subfigure}
\begin{subfigure}[h]{0.4\textwidth}
	\centering
	\includegraphics[scale=0.2]{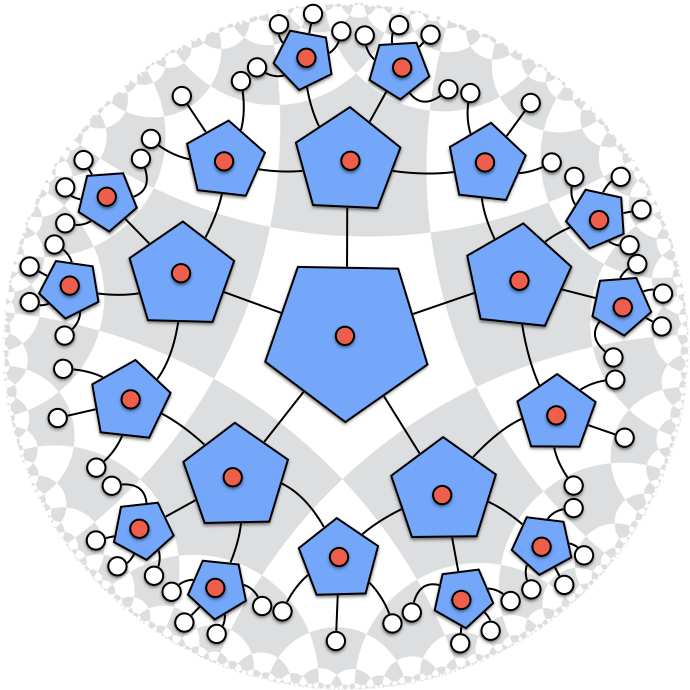}
	\caption{}
	\label{fig:hexagon2}
\end{subfigure}	
\caption{The hyperbolic HaPPY tensor network. (a) depicts the hexagonal state, while (b) depicts the pentagonal code. Credits for these figures go to~\cite{Pastawski:2015qua}.}
\label{fig:HaPPY}
\end{figure} 

The advantages of this network are that it has a constant negative curvature (like the hyperbolic plane), realizes the Ryu-Takayanagi formula for the entanglement entropy of a subsystem and has a nice interpretation as a QECC. The authors have considered two possible networks. The first consists of contracted hexagonal tensors. By starting from a central tensor and then recursively adding layers by contracting each leg with a new tensor, the network of constant negative curvature is built up. At some outer layer, some legs of the tensors are uncontracted. This is the boundary layer and the uncontracted legs are called boundary legs. It represents a state. The state should have the correct entangling features of a state with a holographic dual. Therefore we require that the network satisfies the Ryu-Takayanagi (RT) conjecture for the entanglement entropy. The second network is built from the same hexagonal tensors but now each tensor has one uncontracted or dangling leg that represents a bulk degree of freedom. Because only five legs of each tensor are contracted, this is called the pentagonal network. We will argue that the pentagonal network provides an isometric mapping from physical to logical Hilbert space and diplays QECC.\\

To reproduce RT and QECC, the hexagonal tensors should have special properties. It turns out that it is sufficient to restrict to \textit{perfect tensors}\footnote{For a proof, see~\cite{Pastawski:2015qua}.}. They form a subclass of isometric tensors. A tensor $T:\mathcal{H}_A\rightarrow \mathcal{H}_B$ is \textit{isometric} if it preserves the inner product, i.e. $T^\dagger T = \mathbbm{1}$. The tensor can then be pushed through the isometries
\begin{equation}
TO = TOT^\dagger T = O'T \quad \text{with} \quad O'=T O T^\dagger.
\end{equation}
The indices of the tensor can be explicitly written from $\ket{b} = \sum\limits_a T_{ab}\ket{a}$ with $\ket{a}\in \mathcal{H}_A$ and $\ket{b}\in\mathcal{H}_B$.\\

A \textit{perfect tensor} is a $2n$ tensor with the property that any bipartition of its indices into $A$ and $A^c$ with $\left|A\right|\leq \left|A^c\right|$, T is proportional to an isometric tensor from $A\rightarrow A^c$.\\

A nice property of perfect tensors is that a network of perfect tensors is again a perfect tensor that (when contracted on a basis) describes a state of $2n$ spins such that any set of $n$ spins is maximally entangled with its complement. The proof of this is quite simple. Bipartition the set of spins into $2$ sets of $n$ spins labeled by $A$ en $B$. The reduced density matrix is
\begin{align}
\rho_A &= \text{Tr}_B \rho_{AB} = \sum\limits_{a,b,a'} T_{ab}\ket{a}\bra{a'} T^{\*}_{ba'},\nonumber\\
&= \sum\limits_{a} \ket{a}\bra{a},
\end{align}
where in the second line we have used the isometry property of the perfect tensors. Indeed we find a maximally entangled reduced density matrix. This property ensures that a discrete version of the RT formula holds~\cite{Pastawski:2015qua}.\\

The QECC properties cannot be shown in the hexagonal network because it depicts a state and not a code. But the pentagonal network does. The tensor network can be interpreted as a linear map from $2n-1$ physical spins to a single logical spin. Because of the maximal entanglement between any set of $n$ spins the code will be protected against erasure of any $n-1$ spins, which is just less than half the system size. Like in the $3$-qutrit model the maximal entanglement among physical spins ensures the quantum error correcting properties.\\
The QEC properties of logical operators are relatively easy to illustrate. Attach an operator to one of the dangling bulk legs. This means we act with an operator on a logical spin. If there exists an isometric mapping from the dangling leg to a set of spins on the boundary contained in region $A$, then the operator is reconstructable from the boundary operators in $A$. The isometric map can be constructed by pushing the bulk operator succesfully through a sequence of tensors. Because every tensor is perfect any $3$ incoming legs will be isometrically mapped to three outcoming legs. This sequence of isometries combines into a full isometric map on the network which determines the bulk-boundary map. Different isometries imply different CFT operator constructions which all have the same effect on the code subspace. From that point of view, different reconstructions give rise to the same bulk operator. In a QECC language, one could say that the code is protected against erasure of a set of physical spins if there exists a subalgebra of logical operations which has support only on the set of unerased physical spins.\\

Since the HaPPY network proposal, many other possible holographic codes have been considered. The HaPPY network was extended by considering more general hyperbolic tilings of AdS based on a Coxeter group construction~\cite{Peach:2017npp,Bhattacharyya:2016hbx}. This also allows of building more general backgrounds such as black hole backgrounds or conical defect backgrounds. The point is that these states are obtained by orbifolding the AdS-spacetime. By tesselating AdS and consequently orbifolding the tessellation, a tensor network representing these states can be built. Networks can also be built without using perfect tensors. The literature contains examples of networks based on \textit{pluperfect tensors}~\cite{Yang:2015uoa} and \textit{random tensors}~\cite{Hayden:2016cfa}.

\section{Gauge invariance \& bulk reconstruction}\label{sec:gauginv}
The QEC properties of AdS/CFT rely on seemingly quite crude toy models such as tensor networks. It would be nice if the QEC properties follow in a more natural way. In this section, we will argue that theories with internal gauge symmetry might also display QEC~\cite{Mintun:2015qda}. This provides a more natural interpretation of the quantum error correcting properties of holographic theories because they typically have gauge symmetries. Ideally we would like to apply the argument on actual holographic CFTs but here we will restrict ourselves to a toy lattice model with $O(N)$ invariance.\\

Another symmetry that is very important in the context of bulk reconstruction is diffeomorphism symmetry. The operators constructed from HKLL are not diffeomorphism invariant. Although we've shown in section~\ref{sec:QEC} that a Rindler type of reconstruction in any causal wedge exists, independent of the coordinate system, this is an existence proof and does not show how operators are made diffeomorphism invariant. In this section we will describe a procedure that makes operators explicitly diffeomorphism invariant~\cite{Dirac:1955uv,Donnelly:2015hta,Donnelly:2015taa}. The procedure is called gravitational dressing.

\subsection{QECC and gauge invariance}
\begin{figure}[h]
\centering
\includegraphics[scale=0.4]{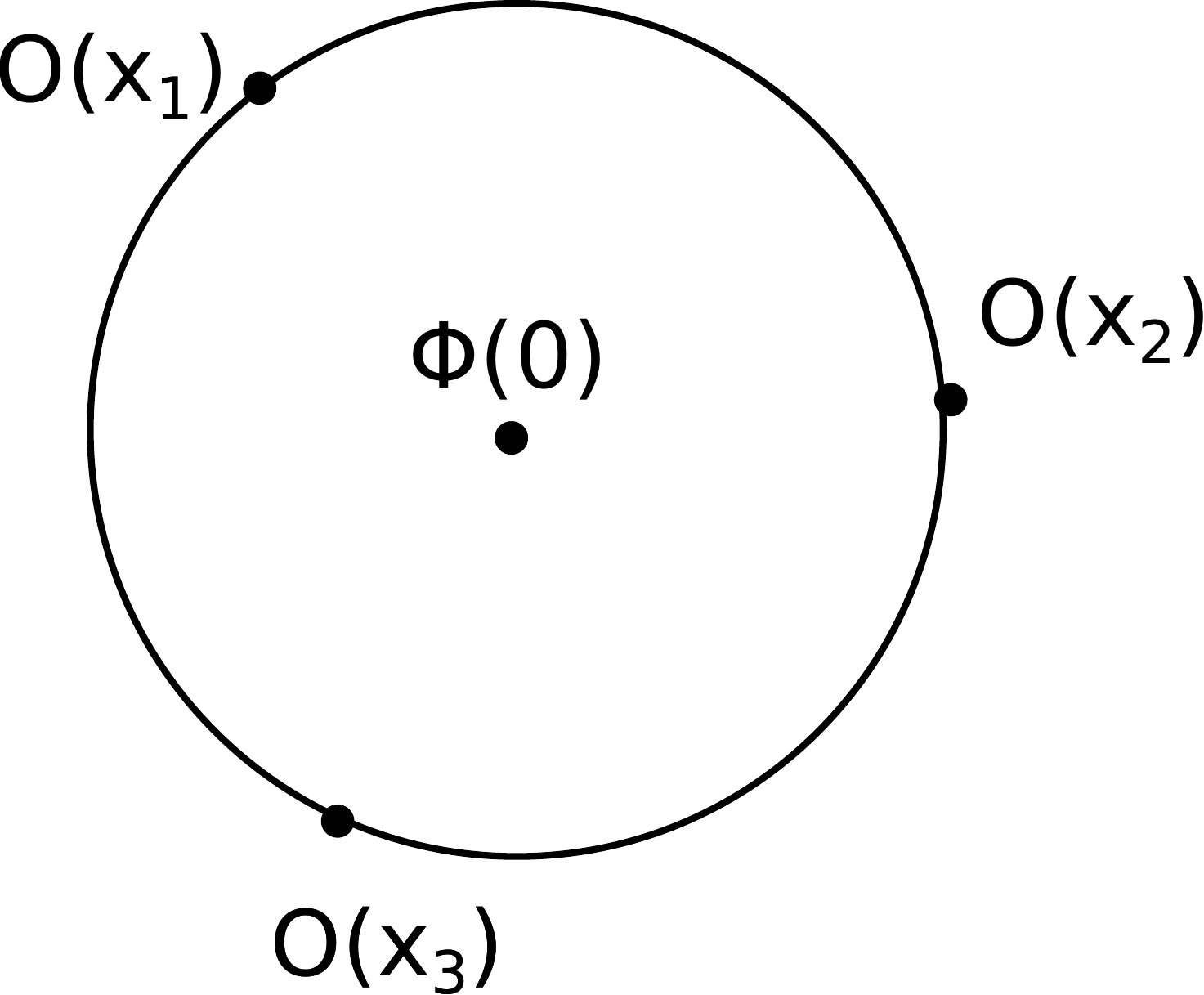}
\caption{Toy scalar field model with $O(N)$ symmetry. The boundary contains $3$ sites, the bulk a single one.}
\label{fig:ONsymmetry}
\end{figure}

Take a scalar field theory on a $1d$ periodic lattice with $3$ points as in figure~\ref{fig:ONsymmetry}. Now endow the theory with an internal $O(N)$ gauge symmetry. The canonical variables in the fundamental representation of $O(N)$ are $(\phi^i_a,\pi^i_a)$ with $a=1,2,3$ denoting the lattice position and $i=1,\ldots,N$ the $O(N)$-index. A complete set of gauge invariant operators on a single site $a$ can be generated by $( \phi_a\cdot \phi_a, \phi_a\cdot\pi_a,\pi_a\cdot\pi_a)$ where we are summing over  the $O(N)$ index. The generators on a single site $a$ form an $SL(2,\mathbbm{R})$ algebra.
\begin{align}
\LT \phi_a\cdot\phi_a, \pi_a\cdot\pi_a\RT &= 4i\phi_a\cdot\pi_a,\\
\LT \phi_a\cdot\pi_a, \pi_a\cdot\pi_a\RT &= 2i\pi_a\cdot\pi_a,\\
\LT \phi_a\cdot\phi_a, \phi_a\cdot\pi_a\RT &= 2i\phi_a\cdot\phi_a,
\end{align}
which by defining $X = \frac{\phi\cdot\phi}{2}, Y= \frac{\pi\cdot\pi}{2}$ and $H= i\phi\cdot\pi$ can be turned into the standard $SL \LF 2,\mathbbm{R}\RF$ algebra
\begin{equation}
\LT X,Y\RT = H \qquad \LT H,Y\RT = -2Y \qquad \LT H,X\RT = 2X.
\end{equation}
The $SL(2,\mathbbm{R})$ allows the construction of an operator that is not gauge invariant but is a singlet with respect to the $SL(2,\mathbbm{R})$ algebra, namely
\begin{equation}
L^{ij}_a = \phi^i_a\pi^j_a - \phi^j_a\pi^i_a.
\end{equation}
Indeed, it is a singlet because it commutes with all generators. We check one commutator explictly
\begin{equation}
\LT L^{ij}_a,\phi_a\cdot\phi_a\RT = 2\LF \phi^k_a \phi^i_a \delta_{kj} - \phi^k_a\phi^j_a\delta_{ik}\RF = 0
\end{equation}
and likewise for the other two commutators. Note that $L^{ij}_a$ is the generator of the $O(N)$ transformations on a single site $a$. The total $O(N)$ generator is $L^{ij} = \sum\limits_a L^{ij}_a$. It annihilates all gauge invariant states, i.e. $L^{ij}\ket{\Psi}=0$. The operator $P_{ab} = L^{ij}_a L^{ij}_b$  then is a gauge invariant operator that commutes with all gauge invariant operators on a single site. So it commutes with all local gauge invariant operators and belongs to the center of the single site operator algebra. Moreover, the operator is $SL(2,\mathbbm{R})$ invariant. By definition of the $P_{ab}$ and of $L^{ij}$ we have that $P_{12} = -P_{22}-P_{32} + L^{ij}L^{ij}_2$. $P_{12},P_{23}$ and $P_{13}$ are clearly different elements of the operator algebra because they act on different boundary sites, but when acting on the space of gauge invariant states this difference disappears. because $L^{ij}$ annihilates the physical states we get that 
$P_{12}\ket{\Psi} = -P_{22}\ket{\Psi} - P_{23}\ket{\Psi}$. An operator involving site $1$ is completely equivalent to an operator only involving sites $2$ and $3$. Because of boundary translation invariance this argument also holds for operators not involving sites $2$ or $3$.  The physical space cannot distinguish between different sites and is robust against erasing one of the sites.\\

An operator that we could interpret as a local bulk operator then is $\Phi = P_{12}+P_{23}+P_{13}$. It is completely gauge invariant and invariant as well under the emerging $SL(2,\mathbbm{R})$ symmetry. Note that this is a quite crude model since we only have a single bulk point.\\

Concludingly, the operators $P_{ab}$ commute with operators in the ``code space" which here is taken to be the space of gauge invariant states, so when the CFT has an internal gauge symmetry, then the corresponding bulk operators automatically display error correcting properties. Note that we have restricted the bulk operators to be also $SL(2,\mathbbm{R})$ invariant. The link between quantum error correction and gauge invariance has been further investigated in~\cite{deBoer:2016yiz,Freivogel:2016zsb}.\\

\subsection{Gravitational dressing}
Well defined operators in a gravitational theory should be diffeomorphism invariant. We will show here that it is impossible to construct operators which are local and at the same time diffeomorphism invariant. In order to construct invariant operators we will start from local operators and dress them gravitationally, which is a non-unique and non-local operation.\\

Take a scalar field in the bulk. Indeed it transforms under an infinitesimal coordinate transformation $x'^\mu=x^\mu-\epsilon \xi^\mu$ as
\begin{equation}
\delta \phi(x) = \phi(x-\epsilon\xi) - \phi(x) = -\epsilon \xi^\mu\partial_\mu \phi + O(\epsilon^2).
\end{equation}
We can create a non-local operator $\Phi_V (x) = \phi\LF x^\mu + V^\mu(x)\RF $. Clearly if $V$ transforms as $\delta V^\mu = \epsilon \xi^\mu$ then $\Phi_V$ is diffeomorphism invariant. For small values of the vector field $\Phi_V$ can be written as
\begin{equation}
\Phi_V(x) = \phi(x) + V^\mu \partial_\mu \phi + O(V^2),
\end{equation}
which is an infinitesimal version of a non-local operator acting on $\Phi_V$, namely
\begin{equation}
\Phi_V = e^{i V^\mu P_\mu} \phi(x) e^{-iV^\mu P_\mu}
\label{eq:PhiDressed}
\end{equation}
with $P_\mu= -i\partial_\mu$ the momentum operator. The non-locality of $\Phi_V$ follows from the infinite number of derivatives that are present in its definition (\ref{eq:PhiDressed}). The vector field that we have defined here is precisely analagous to the gauge field in a $U(1)$ theory. In that case the gauge field causes an electromagnetic dressing~\cite{Dirac:1955uv}. Here the vector field is the gauge field associated to diffeomorphisms and the associated dressing of the scalar operator is called \textit{gravitational dressing}~\cite{Donnelly:2015hta}.\\

We need to provide a sensible solution for the dressing field $V$ such that it transforms precisely like $\delta V^\mu = \epsilon \xi^\mu$. We are only going to be interested in field solutions to $O(\epsilon)$ which is the linearized level, and since $V^\mu$ is a gravitational object it should be formed from linear combinations of the metric perturbations $h_{\mu\nu}$. For simplicity we take the background metric to be the flat metric here. The proposed ansatz is
\begin{equation}
V^{\mu} (x) = \epsilon \int d^4 x' f^{\mu\nu\lambda}(x,x') h_{\nu\lambda}(x'),
\end{equation}
with $f$ symmetric in the last two indices because $h$ is symmetric as well. Demanding that $V$ transforms in the correct way, puts constraints on $f^{\mu\nu\lambda}$. Under a diffeomorphism
\begin{align}
\delta V^{\mu} &= \epsilon\int d^4 x' f^{\mu\nu\lambda}(x,x') \delta h_{\nu\lambda}(x'),\\
&= - \epsilon \int d^4 x' f^{\mu\nu\lambda}(x,x')\LF \partial_\nu \xi_\lambda(x') + \partial_\lambda \xi_\nu(x')\RF,\\
&= -2\epsilon \int d^4 x' f^{\mu\nu\lambda}(x,x') \partial_\nu \xi_\lambda (x'),\\
&= 2\epsilon \int d^4 x' \partial_\nu^{'} f^{\mu\nu\lambda}(x,x') \xi_\lambda (x'),
\end{align}
where in the second line we have used the rule for transformations of the metric $\delta h_{\mu\nu} = -\partial_\mu \xi_\nu - \partial_\nu \xi_\mu$ and in the last line we have integrated by parts. From demanding $\delta V^\mu= \epsilon \xi^\mu$, it follows that $f$ must satisfy
\begin{equation}
\partial^{'}_\nu f^{\mu\nu\lambda} (x,x') = \frac{1}{2}\delta^{(4)}(x-x') \eta^{\mu \lambda}.
\label{eq:constraint}
\end{equation}
Solutions to this equation will provide sensible dressings.\\

In electromagnetism an electron can be dressed with a Wilson line, but also with an entire electric field. Both procedures are valid and give different non-local operators. In the gravitational context various similar types of dressing are possible. Here we will focus on one particular type called gravitational Wilson line dressing. A gravitational  Wilson line is attached to an operator and extends all the way to infinity. Such a dressing has a divergent energy density and will therefore actually be unstable, and decay to a more symmetric type of dressing like Coulomb dressing, but from the point of view of diffeomorphism invariance the dressing field ensures the scalar field to be invariant.\\

To define the Wilson line, one needs to pick a platform at infinity and a point on this platform to which the Wilson line is attached. Coordinates $x^\mu = (\tilde{x}^\mu,\tilde{z})$ are chosen such that $\tilde{z}$ is the coordinate orthogonal to the platform while the $\tilde{x}^\mu$ are the coordinates on the platform. These are the normal coordinates in which $\tilde{z}$ actually parametrizes the geodesic distance to the platform. If we fix the axial gauge
\begin{equation}
h_{\tilde{z}\tilde{\mu}}=0,
\end{equation}
then we can define the Wilson line to be trivial in this coordinate set, i.e. $\Phi(\tilde{x}^\mu,\tilde{z}) = \phi(\tilde{x}^\mu,\tilde{z})$. We would now like to derive how the Wilson line looks like in a general different coordinate set. Perform a diffeomorphism, such that we turn on a vector field and to new coordinates which to linear order are defined by $\tilde{x}^\mu = x^\mu - V^\mu(z,x)$ and $\tilde{z} = z - V^z (z,x)$. The scalar operator gets non-locally dressed by the vector field as in~(\ref{eq:PhiDressed}). The vector field is determined by solving the constraint equation~(\ref{eq:constraint}). Appropriate solutions for $f$ are~\cite{Donnelly:2015hta}
\begin{equation}
\left\{
\begin{matrix}
f^{z\nu\lambda} = \frac{1}{2} \eta^{\nu z}\eta^{\lambda z} \int\limits_0^{+\infty} ds \delta^{(4)}(x'-x-s\hat{z}),\\
f^{\tilde{\mu}\nu\lambda} = \int\limits_0^{+\infty} ds \eta^{z\nu} \eta^{\tilde{\mu}\lambda} \delta^{(4)}(x'-x-s\hat{z}) - \frac{1}{2}\int\limits_0^{+\infty} ds \int\limits_s^{+\infty} ds' \eta^{z\nu}\eta^{\tilde{\mu}\lambda} \partial_{\tilde{\mu}} \delta^{(4)}(x'-x-s\hat{z}).
\end{matrix}
\right.
\end{equation}
We will explicitly check the first component.
\begin{align}
\partial^{'}_\nu f^{z\nu\lambda}(x,x') &= \frac{1}{2}\int\limits_{0}^s ds \eta^{\lambda z} \partial_z \delta^{(4)}(x'-x-s\hat{z}),\nonumber\\
&= \frac{1}{2} \int\limits_{0}^{+\infty} ds \eta^{\lambda z} \partial_s \delta^{(4)}(x'-x-s\hat{z}),\nonumber\\
&= \frac{1}{2}\eta^{\lambda z}\delta^{(4)}(x'-x).
\end{align}
A residual freedom is contained in $f^{\mu\nu\lambda}$ in the sense that solutions to~(\ref{eq:constraint}) can be added to its definition if they vanish when integrated against $h_{\nu\lambda}(x')$. The vector field $V^\mu$ that satisfies the ansatz with the $f^{\mu\nu\lambda}$ as above will be the dressing field that turns the non-gauge invariant local bulk scalar operator into a non-local gauge invariant scalar operator.\\ The bottom line of gravitational dressing is very simple. In a gauge invariant theory, it is impossible to create a local bulk excitation without also exciting its associated gravitational field. Other types of dressing are also possible like for example Coulomb dressing which creates a Coulomb field that uniformly spreads and is attached to a pointlike charge.
\section{Bulk operators in $AdS_3/CFT_2$ create Virasoro cross-cap states}\label{sec:crcap}
\subsection{Cross-cap states}
Because of the absence of local gravitational degrees of freedom it is simpler to look at bulk reconstruction in $AdS_3/CFT_2$~\cite{Nakayama:2015mva,Nakayama:2016xvw,Verlinde:2015qfa,Lewkowycz:2016ukf}. In this context another approach to bulk reconstruction is possible. The advantage of this new approach will be that it is able to reconstruct bulk operators non-perturbatively, intrinsic to the CFT (without referral to equations of motion), in a background independent and diffeomorphism invariant way. The disadvantage is that the bulk operators change the topology of the CFT worldsheet and therefore act inherently non-local in the CFT.\\
\begin{figure}[h]
\centering
\includegraphics[scale=0.4]{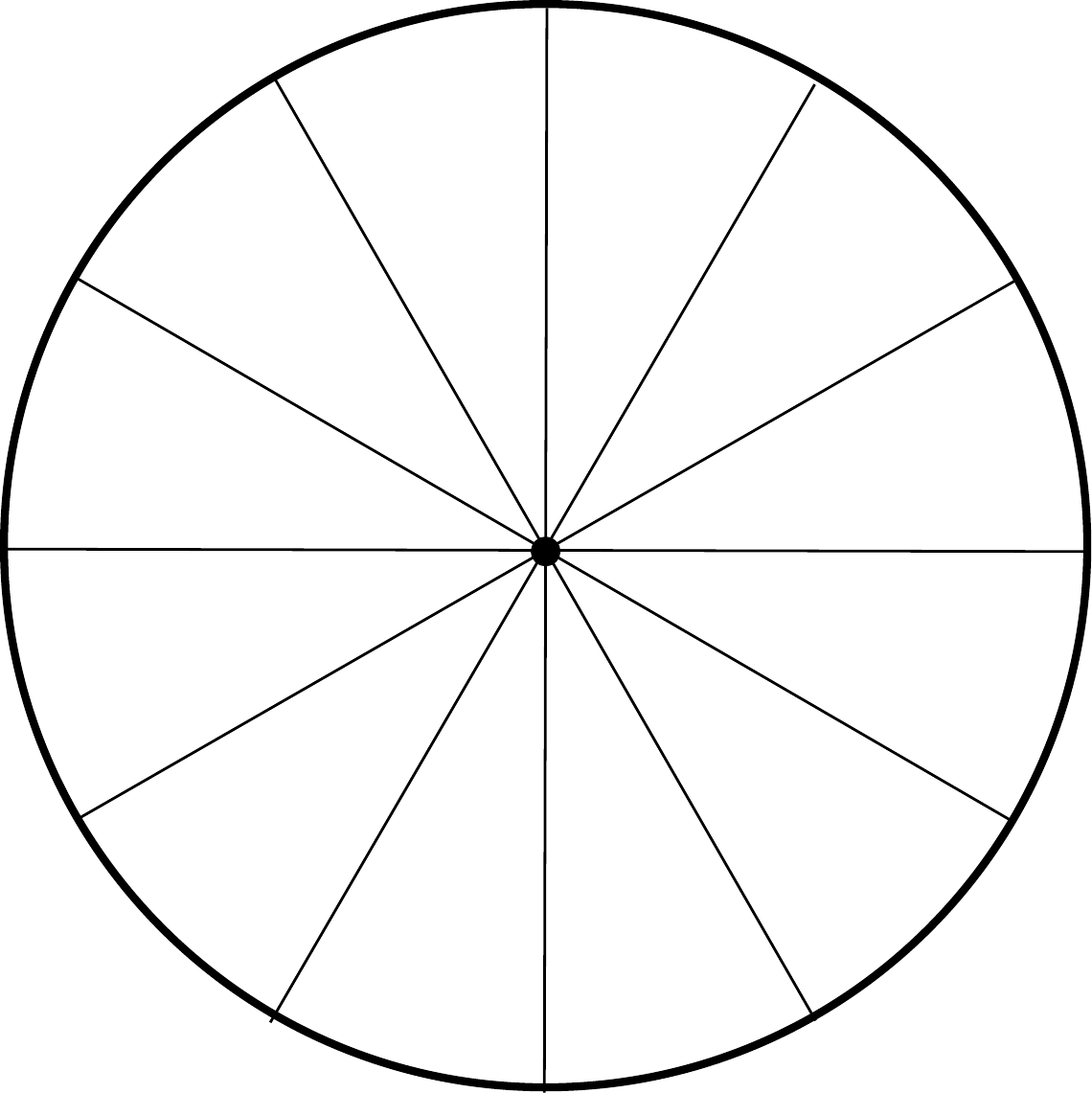}
\caption{Operator in the center of AdS (black dot) created from the intersection of radial geodesics.}
\label{fig:OpCenter} 
\end{figure}
We will work in Euclidean signature here\footnote{In static states the Wick rotation from the Lorentzian CFT is straightforward. Dynamic situations are subtler as singularities in the complex plane might have to be avoided as one analytically continues. We will not bother with this complication and restrict to the Euclidean $AdS_3/CFT_2$ correspondence.}. A bulk operator is defined on a point in AdS, so the first question is how to characterize a point in a coordinate invariant way. One way to single out a point is to view it as the intersection of a class of boundary anchored geodesics. The operator in the center is the easiest to consider. It is simply the intersection of radial geodesics. This situation is drawn in figure~\ref{fig:OpCenter} and although it depicts just a $2d$ slice of AdS, the center is actually constructed from all radial geodesics on the $S^2$ boundary. This construction is invariant under swapping the endpoints of the geodesics $\varphi \rightarrow \varphi +\pi$ or in holomorphic Poincar\'e coordinates $z\rightarrow \frac{-1}{\bar{z}}$. A local bulk operator in the center of AdS must be invariant under this symmetry transformation. This symmetry in fact implies that bulk operators $\Phi(X)\equiv\Phi(z,y)$  are not defined as CFT operators on the $S^2$ but rather on $S^2/\mathbbm{Z}_2=\mathbbm{R}\mathbbm{P}_2$. The transformation $z'= -\frac{1}{\bar{z}}$ is the complex conjugate of a M\"obius transformation which is part of the conformal group and the generator of those is the stress tensor $T(z)$. By assumption of radial quantization $\Phi(X)\ket{0} = \ket{\Phi(X)}$ and the operator at the center $\Phi(0)$ creates a state $\ket{\Phi}$. The infinitesimal invariance condition becomes 
\begin{equation}
\LF T'(z') - T(z)\RF \ket{\Phi} = 0.
\label{eq:CCcond}
\end{equation}
In Virasoro modes,
\begin{align}
T(z) &= \sum\limits_n L_n z^n,\nonumber\\
T'(z') &= \bar{T}\LF \frac{-1}{\bar{z}}\RF = \sum\limits_n \bar{L}_n \LF \frac{-1}{z}\RF^n,\nonumber\\
&= \sum\limits_n \bar{L}_{-n}  (-1)^n z^n.
\end{align}
The invariance condition in modes therefore reads
\begin{equation}
\LF L_n - (-1)^n\bar{L}_{-n}\RF \ket{\Phi} = 0 .
\end{equation}
This is the defining equation of a Virasoro cross-cap state~\cite{Blumenhagen:2009zz}. So a bulk operator at the center of $AdS_3$ creates a Virasoro cross-cap state. Even though we started from a bulk construction in terms of geodesic intersections, our final definition of bulk operators is completely intrinsic to the CFT without any reference to the bulk.

Of course bulk operators should be defined not only at the center but at any bulk point. The isometry invariance of $AdS_3$ allows to pick an isometry transformation to move the center to an arbitrary bulk point. The isometry group of $AdS_3$ is $SL(2,\mathbbm{R})$ and corresponds to the M\"obius group of the CFT. An element of this group can be parametrized as
\begin{equation}
g(X) = e^{-iHt} e^{\varphi l_0} e^{\frac{\rho}{2}\LF l_{-1} - l_{1}\RF},
\end{equation}
with
\begin{equation}
H = L_0 + \bar{L}_0 \qquad l_0 = L_0- \bar{L_0} \qquad l_{\pm 1} = L_{\pm 1} - \bar{L}_{\mp 1}.
\end{equation}
From the bulk point of view $(t,\varphi,\rho)$ are interpreted as the global coordinates on AdS but from the CFT point of view they are just labels parameterizing the transformation. Applying the element to a state with a bulk operator at the center of AdS, creates a state with a bulk operator at the point $X=(t,\varphi,\rho)$, i.e. $g(X)\ket{\Phi} = \ket{\Phi(X)}$, which is again a Virasoro cross-cap state where the cross-cap has parameters $(t,\varphi,\rho)=(y,x^\mu)$ with the first are parameters corresponding to global AdS coordinates, whereas the second are the corresponding Poincar\'e coordinates. The M\"obius transformation changes the identification of points from $z\rightarrow -\frac{1}{\bar{z}}$ to 
\begin{equation}
z\rightarrow z' = x - \frac{y^2}{\bar{z}-\bar{x}}.
\end{equation}
The Hilbert space of states of the $CFT_2$ contains many superselection sectors, labeled by the highest weight states, and each sector contains a solution to~(\ref{eq:CCcond}). A cross-cap state is therefore not just labeled by $X$ but also by the sector $h$ it belongs to. Each sector is in one-to-one correspondence with a primary operator so the labels $h$ can be taken to be the conformal weight of the operator $O_h$ that is dual to the local bulk operator. Said otherwise, a cross-cap doesn't just put geometric identifications on the CFT worldsheet but also projects onto an irreducible representation of the Hilbert space. It can then also be shown that the cross-cap operator $\Phi(X)$ with boundary conditions $\Phi(x,y\rightarrow 0) = y^{-2h} O_h(x)$ solves the wave equation $(\Box - m^2)\Phi = 0$ to leading order in $\frac{1}{N}$~\cite{Nakayama:2015mva,Lewkowycz:2016ukf}. \\

\subsection{Gravitational dressing}

Say that in a coordinate system $\chi = (\tilde{y},\tilde{x}^\mu)$ the dressing is trivial. Then analogous to previous discussions a gauge choice can be made such that the bulk operator is a global cross-cap state $\Phi^{(0)}(\chi)$. The Virasoro modes with $n\geq 2$ generate diffeomorphisms in the bulk.  But $\Phi^{(0)}$ is not a diffeomorphism invariant operator. Say that we consider a diffeomorphism $X=\chi(X) + \xi(X)$. In order to make the operator diffeomorphism invariant we will need to turn on gravitational dressing. The dressing turns $\Phi^{(0)}$ into a Virasoro\footnote{ We call it Virasoro because the Virasoro group is the group of non-vanishing diffeomorphisms on the boundary $\text{Diff}(S^1) \times \text{Diff}(S^1)$.} cross-cap operator
\begin{equation}
\Phi(X) = e^{\mathcal{V}(X)}\Phi^{(0)}(X) e^{\mathcal{V}^\dagger (X)}.
\end{equation}
After performing radial quantization, the action of the dressing is
\begin{equation}
\ket{\Phi(X)} = e^{\mathcal{V}(X)} \ket{\Phi^{(0)}(X)},
\end{equation}
In the particular case of projecting onto the irreducible representation with $h=0$ , the map becomes
\begin{equation}
\ket{\mathbbm{1}(X)} = e^{\mathcal{V}(X)}\ket{0}.
\end{equation}
So even the identity cross-cap operator is a non-trivial operator, namely it is precisely the gravitational dressing operator. To find a specific form for the dressing let's focus on the identity operator inserted at the origin of $AdS_3$. The Taylor series expansion of the identity is
\begin{equation}
\ket{\mathbbm{1}(0)} = e^{\mathcal{V}}\ket{0} = \ket{0} + \mathcal{V}\ket{0} + \ldots,
\end{equation}
and because it is a cross-cap operator it necessarily satisfies
\begin{equation}
\LF L_n - (-1)^n \bar{L}_{-n}\RF \ket{\mathbbm{1}(0)} =0.
\label{eq:constraint1}
\end{equation}
In~\cite{Lewkowycz:2016ukf} it is shown by making use of the Virasoro algebra that the constraint is met by the following dressing operator
\begin{equation}
\mathcal{V} = \sum\limits_{n=2}^{+\infty} \frac{(-1)^n 12L_{-n}\bar{L}_{-n}}{c( n^3-n)}+\ldots,
\end{equation}
where the dots indicate subleading terms in a $1/c$ expansion.

\subsection{Bulk-boundary correlator}
As a last example of this set-up we show that the interpretation of bulk operators in $AdS_3$ as Virasoro cross-cap operators, recovers correct correlation functions. It can be shown that the bulk-bulk two point function seen as the overlap of two Virasoro cross-cap states equals the AdS-propagator~\cite{Verlinde:2015qfa,Lewkowycz:2016ukf}, but here we will focus on the slightly simpler example of the two-point function of one bulk operator $\Phi(X)$ and one operator $O_h(x)$. The two-point function can only be non-zero if the two belong to the same irreducible representation of the Hilbert space and thus $\Phi(x,y\rightarrow 0)= \lim\limits_{y\rightarrow 0} y^{-2h}O_h(x)$. We study
\begin{equation}
\langle O_h(x_1,\bar{x}_1) \Phi (y, x_2,\bar{x}_2)\rangle.
\end{equation}
The cross-cap operator creates an identification $\mathbb{R}\mathbbm{P}_2 = S^2/\mathbbm{Z}_2$. So our correlation function is just another way of studying the one-point function $\langle O_h\rangle_{\mathbb{R}\mathbbm{P}_2 }$ on $\mathbbm{R}\mathbbm{P}_2$. The $\mathbbm{Z}_2$-identification can be undone by doubling the worldsheet of the CFT. This is called the Schottky double. Since the Schottky double is obtained by just doubling the worldsheet with the operator insertion on it, on the Schottky double we will have two local operator insertions, namely one on the original position and one on the image point under the $\mathbbm{Z}_2$ transformation. So we have mapped the correlation function to a standard two point function of local operators
\begin{equation}
\langle O(x_1,\bar{x}_1) O(x'_1,\bar{x}^{'}_1)\rangle \qquad \text{with} \quad x'_1 = x_2 - \frac{y^2}{\bar{x}_1 - \bar{x}_2}.
\end{equation}
With the correct Jacobian factor $\LF \frac{\partial z'}{\partial z}\RF^h=\frac{y^{2h}}{(\bar{x}_1-\bar{x}_2)^{2h}}$ where we have only acted on the left moving components~\cite{Lewkowycz:2016ukf}, we arrive at 
\begin{equation}
\langle O_h(x_1,\bar{x}_1) \Phi_h (y, x_2,\bar{x}_2)\rangle = \frac{y^{2h}}{(y^2+ \left| x_{12}\right|^2)^{2h}}
\end{equation}
which agrees precisely with the bulk-boundary propagator in AdS.

\section{Conclusion}\label{sec:conclusion}
The holographic correspondence predicts an isomorphism between bulk and boundary Hilbert spaces and between bulk and boundary operator algebras. The representation of bulk operators is obscured in the holographic dictionary and local bulk operators are non-locally smeared over the boundary. This map can be explicitly computed via the HKLL reconstruction method which involves convoluting the boundary operators with a smearing function which in some sense is a propagator. The HKLL method has various shortcomings that we have addressed in these notes.
\begin{itemize}
\item The smearing function is not unique. It can be shifted by terms which vanish when integrated on a CFT operator distribution. 
\item To obtain the smearing function, the bulk equations of motion need to be solved. This prevents us from having a definition of bulk operators completely intrinsic to the CFT. 
\item The bulk operators in this way are not diffeomorphism invariant.
\item The operator reconstruction depends on the choice of reconstruction wedge.
\item Because we have solved the bulk equations of motion in a particular background, the  method is background dependent.
\item HKLL reconstruction is perturbative in $1/N$, whereas the operator algebra isomorphism should hold non-perturbatively.
\end{itemize}
The dependence on choice of wedge has led to the discovery of the quantum error correcting properties of AdS, which has led us to discussing tensor network toy models for AdS/CFT. Quantum error correcting properties seem to be very surprising at first, but we have shown that similar properties can arise automatically in theories with an internal gauge symmetry. A model of a scalar field on a lattice with $3$ sites and with a local internal $O(N)$ symmetry  has the property that operators with support on different sites are equivalent when evaluated on the space of gauge invariant states. The only well-defined operator algebra in a gravitational theory consists of diffeomorphism invariant operators and for that reason local bulk operators need to be gravitationally dressed, for example by a gravitational Wilson line. At the same time the gravitational dressing actually turns the local operator into a non-local one. Just like in electromagnetism, local gauge invariant operators do not really exist, they are always non-local to some extent. Of course, in a particular gauge, the dressing can be trivial. Finally, we have constructed bulk operators which are diffeomorphism invariant, non-perturbatively well defined, background independent in a theory of pure 3d gravity with negative cosmological constant. The bulk operators are operators that create Virasoro cross-cap states. We have seen that these reproduce the correct propagator. The cross-cap operators are intrinsically non-local quantities and are gravitationally dressed.\\

A couple of important developments concerning reconstruction of the bulk haven't been discussed here. Very recently, bulk reconstruction has been considered using concepts like \textit{modular hamiltonian} and \textit{modular flow}\cite{Faulkner:2017vdd}. The modular hamiltonian is defined as the operator $-\log \rho$ where $\rho$ is the density matrix under consideration and modular flow is the flow generated by the modular Hamiltonian. These concepts have shown to provide a useful language to talk about bulk reconstruction.\\

Secondly, the RT formula points to an important link between geometry and entanglement. By now a vast amount of literature exists, that does not just try to reconstruct bulk operators but tries to explain the emergence of the bulk spacetime from entanglement in the CFT. This would provide a stronger paradigm where the AdS/CFT doesn't describe  just a duality but where the bulk actually emerges from the CFT. This program is to be continued.\\

Many of the problems with HKLL have been partially resolved, but many of the resolutions come with their own problems. For example the tensor network construction is a geometric representation of a state at a fixed time. So far it is unclear how to build a spacetime from this. The cross-cap construction on the other hand only works for geometries which are pure 3d gravity solutions without matter. Inserting matter into the game would make the spacetime dynamical and it remains a question whether the cross-cap construction still works in that case. It would be interesting to understand better the link between gauge invariance and the emergent quantum error correcting properties of AdS/CFT. Some work has been done by e.g. Freivogel and de Boer~\cite{deBoer:2016yiz,Freivogel:2016zsb} using BRST symmetry, but a complete understanding hasn't been reached yet. Another problem is that we do not know the size of the code subspace. We think of it as some low energy effective subspace where bulk effective field theory holds, but as yet there is no precise definition of the code subspace in AdS/CFT. The AdS/CFT correspondence in its most precise sense emerges from string theory. Does there exist a string theory embedding in which these ideas concerning bulk reconstruction can be illustrated and made more concrete? Hopefully a next series of Modave lectures can answer these questions.

\section*{Acknowledgments}
The author would like to thank the co-organizers of the XIII Modave School in Mathematical Physics for the opportunity of presenting this series of lectures. The author would especially like to thank G. S$\acute{\text{a}}$rosi for various discussions on the subject and the careful reading of the manuscript. The author is also grateful towards the authors of~\cite{Pastawski:2015qua} for allowing to use the nice figures of their paper. T.D.J is Aspirant FWO Vlaanderen.

\begin{appendix}

\section{Scalar field in AdS/CFT}
\label{sec:app0}
The AdS/CFT duality can be phrased as the equivalence of the CFT and gravitational path integral
\begin{equation}
Z_{\text{grav}}\LT O\RT = Z_{CFT} \LT \phi_0\RT= \langle e^{\int\limits d^d x \phi_0(x) O(x)}\rangle_{\text{CFT}}.
\end{equation} 
Semiclassically, the gravitational path integral will localize on saddle point solutions, so we get to an equivalence
\begin{equation}
Z_{CFT}\LT O\RT = \sum\limits_{i} e^{-S_{\text{grav},i}}\approx e^{-S_{\text{grav},0}}.
\end{equation} 
with $S_{\text{grav},i}$ the action of the $i$'th gravitational saddle and the last approximation is done selecting the dominant saddle. The gravitational action should be evaluated at the boundary. Since the boundary is asymptotic we introduce a cutoff at $y=\epsilon$ and evaluate the action at this cutoff surface. As $\epsilon\rightarrow 0$ the gravitational action is divergent which means we will have to renormalize it. After subtracting the correct counterterms to cancel the divergences we arrive at a well defined action $S^{\text{ren}}_{\text{grav}}$. Standard functional methods tell us that the expectation value of the operator $O$ can be computed as
\begin{equation}
\langle O(x) \rangle_{\text{CFT}} = \frac{ \delta \log Z_{\text{CFT}}}{\delta \phi_0(x)} = \left. y^{d-\Delta}\frac{\delta S^{\text{ren}}_{\text{grav}}}{\delta \phi(x,y)}\right|_{y=\epsilon}.
\label{eq:OExpect}
\end{equation}
To determine the action at $y=\epsilon$ we just need to plug in the field solution near the boundary~(\ref{eq:scalexpand}). Notice that indeed the action will be diverging and since the first term in~(\ref{eq:scalexpand}) is the only non-vanishing term in the limit $\epsilon\rightarrow 0$ it follows immediately from plugging it into $\langle e^{\int d^d x \phi_0(x) O(x)}\rangle$ that $\phi_0(x)$ indeed agrees with the $\phi_0(x)$ that appeared in the expansion~(\ref{eq:scalexpand}) and therefore the leading mode acts as the source dual to operator $O$. Now we show that the subleading mode equals the expectation value of the operator $O$ in absence of a source $J$. The on-shell gravity action is just a boundary term
\begin{equation}
S = \int\limits_{\partial M} \sqrt{-g} \phi g^{\mu\nu}\partial_\nu\phi n_\mu d^d x.
\end{equation} 
Since the boundary is located at $y=\epsilon$ the normal vector is in the $y$-direction and we find
\begin{equation}
S = \int\limits_{y=\epsilon} \sqrt{-g} \phi g^{yy}\partial_y \phi d^d x.
\end{equation}
Now we plug in~(\ref{eq:scalexpand}) and explicitly write the metric of AdS at $y=\epsilon$ and we get
\begin{align}
S &= \epsilon^{d-2\Delta}(d-\Delta)\int \phi_0^2(x) d^d x + d \int \phi_0(x)\tilde{\phi}(x) d^d x + \epsilon^{2\Delta-d} \Delta \int \tilde{\phi}^2 d^d x + \ldots.
\end{align}
We subtract the pieces that diverge as $\epsilon\rightarrow 0$ and drop subleading terms. In the end a variation of $S$ with respect to the source $\phi_0$ shows that $\langle O \rangle \sim \tilde{\phi}$. The subleading mode equals the sourceless expectation of the dual operator in the CFT. We have only denoted the relation between $\tilde{\phi}$ and the vacuum expectation value of $O$ with a $\sim$ symbol, because in fact the prefactor that one gets from a simple evaluation of the action is not consistent with the Ward identities. The correct prefactor is discussed in~\cite{Freedman:1998tz}.
\section{Poincar\'e smearing function}
\label{sec:app1}
The smearing is non-local in global coordinates. Here we will show that in Poincar\'e coordinates there exists a different smearing function $K^{\text{Poincar\'e}}\neq K^{\text{global}}$ which is local as $\phi\rightarrow \tilde{\phi}$. For simplicity we will work in $\text{AdS}_3$. The scalar field equation can be solved in a mode expansion in Poincar\'e coordinates
\begin{equation}
\phi(t,x,y) = \int\limits_{\omega> \left|k\right|} d\omega dk a_{\omega k } e^{-i\omega t} e^{ikx} y J_{\Delta-1}\LF \sqrt{\omega^2-k^2}y\RF + \text{c.c}
\end{equation}
with $J_{\Delta-1}$ a Bessel function. The mode expansion of the boundary field in Poincar\'e coordinates $\tilde{\phi}^P$ follows directly
\begin{align}
\tilde{\phi}^{P}(t,x,y) &= \lim\limits_{y\rightarrow 0} y^{-\Delta} \phi(t,x,y),\\
&= \frac{2^{1-\Delta}}{\Gamma\LF\Delta\RF}\int\limits_{\omega>\left|k\right|} d\omega dk a_{\omega k} e^{-i\omega t} e^{ikx}\LF \omega^2-k^2\RF^{\frac{\Delta-1}{2}},
\end{align}
where the small argument expansion of the bessel function has been used
\begin{equation}
J_{\Delta-1} (\sqrt{\omega^2-k^2}y) = \frac{ 1}{\Gamma(\Delta)2^{\Delta-1}} y^{\Delta-1} \LF \omega^2-k^2\RF^{\frac{\Delta-1}{2}}.
\end{equation}
Inverting this relation and plugging back into the mode expansion of $\phi$ results into
\begin{equation}
\phi(t,x,y) = 2^{\Delta-1} \Gamma\LF \Delta\RF \int\limits_{\omega>\left|k\right|} d\omega dk e^{-i\omega t}e^{ikx} \frac{y J_{\Delta-1}\LF \sqrt{\omega^2-k^2}y\RF}{\LF \omega^2-k^2\RF^{\frac{\Delta-1}{2}}}\tilde{\phi}^P(\omega,k) + \text{c.c}
\label{eq:scalPoincare}
\end{equation}
where we have also used the Fourier transform of the boundary field. The Poincar\'e smearing function can now be read off:
\begin{equation}
K(\left. t,x,y\right| t',x') = \frac{2^{\Delta-3}\Gamma\LF \Delta\RF y}{\pi^2} \int\limits_{\omega>\left|k\right|} d\omega dk e^{-i\omega (t-t')} e^{ik(x-x')} \frac{ J_{\Delta-1}\LF \sqrt{\omega^2-k^2}y\RF}{\LF \omega^2-k^2\RF^{\frac{\Delta-1}{2}}}
\end{equation}
The integral over momentum space in~(\ref{eq:scalPoincare}) can be worked out to give an ordinary reconstruction in position space. The associated smearing will have support not on the full Poincar\'e patch, but on a patch which becomes smaller as the bulk operator moves towards the origin. To work out the momentum integral we will need to analytically continue the boundary fields on the complex $x$-plane. The following identities on the Bessel functions are needed:
\begin{align}
&J_{\Delta-1}(b) = b^{\Delta-1} \frac{2^{2-\Delta}}{\Gamma \LF \Delta-1\RF} \int\limits_0^1 r dr \LF 1-r^2\RF^{\Delta-2} J_0(br),\\
&J_0\LF r\sqrt{\omega^2-k^2}\RF = \frac{1}{2\pi} \int\limits_0^{2\pi} d\theta e^{-i\omega r \sin\theta - kr\cos\theta}.
\end{align}  
The scalar field is turned into
\begin{equation}
\phi(t,x,y) = \frac{\Delta-1}{\pi}\int\limits_0^1 r dr (1-r^2)^{\Delta-2} y^{\Delta}\int\limits_0^{2\pi} d\theta \int\limits_{\omega > \left|k \right|} d\omega dk e^{-i\omega (t+ry\sin\theta)} e^{ik\LF x+iry\cos\theta\RF}.
\end{equation}
The integrand looks a lot nicer in coordinates
\begin{equation}
\left\{
\begin{matrix}
y' = ry\cos\theta,\\
t' = ry\sin\theta,
\end{matrix}
\right.
\end{equation}
The integration range is a disk of radius $1$, which in the Cartesian coordinates is described by $t'^2+y'^2 \leq y^2$. The integrals turn into
\begin{equation}
\phi(t,x,y)= \frac{\Delta-1}{\pi} \int\limits_{ y^{'2} + t^{'2}\leq y^2} dy' dt' \LF \frac{y^2-y'^2 -t'^2}{y}\RF^{\Delta- 2} \int\limits d\omega dk e^{-i\omega (t-t')} e^{ik (x+iy')} \tilde{\phi}^P(\omega,k).
\end{equation}
The Fourier transform of the boundary field can be recognized and after the transformation the final result reads
\begin{equation}
\phi(t,x,y) = \frac{\Delta-1}{\pi} \int\limits_{ y^{'2} + t^{'2} \leq y^2} dy' dt' \LF \frac{ y^2-y^{'2}-t^{'2}}{y}\RF^{\Delta-2} \tilde{\phi}^P (t+t',x+iy').
\end{equation}
In these coordinates the associated smearing function is very simple but we had to analytically continue the boundary $x$-coordinate into the complex plane. The advantage is that the bulk operator manifestly becomes local as it approaches the boundary.

\end{appendix}

\bibliographystyle{JHEP}
\bibliography{Bibliography}

\providecommand{\href}[2]{#2}\begingroup\raggedright\begin{thebibliography}{10}



\bibitem{Maldacena:1997re}
  J.~M.~Maldacena,
  Int.\ J.\ Theor.\ Phys.\  {\bf 38} (1999) 1113
   [Adv.\ Theor.\ Math.\ Phys.\  {\bf 2} (1998) 231]
  doi:10.1023/A:1026654312961
  [\arXiv{hep-th/9711200}].


\bibitem{Witten:1998qj}
  E.~Witten,
  Adv.\ Theor.\ Math.\ Phys.\  {\bf 2} (1998) 253
  [\arXiv{hep-th/9802150}].

\bibitem{Balasubramanian:1999ri}
  V.~Balasubramanian, S.~B.~Giddings and A.~E.~Lawrence,
  JHEP {\bf 9903} (1999) 001
  doi:10.1088/1126-6708/1999/03/001
  [\arXiv{hep-th/9902052}].

\bibitem{Giddings:1999qu}
  S.~B.~Giddings,
  Phys.\ Rev.\ Lett.\  {\bf 83} (1999) 2707
  doi:10.1103/PhysRevLett.83.2707
  [\arXiv{hep-th/9903048}].


\bibitem{Ryu:2006bv}
  S.~Ryu and T.~Takayanagi,
  Phys.\ Rev.\ Lett.\  {\bf 96} (2006) 181602
  doi:10.1103/PhysRevLett.96.181602
  [\arXiv{hep-th/0603001}].

\bibitem{VanRaamsdonk:2009ar}
  M.~Van Raamsdonk,
  \arXiv{arXiv:0907.2939} [hep-th].

\bibitem{Hamilton:2005ju}
  A.~Hamilton, D.~N.~Kabat, G.~Lifschytz and D.~A.~Lowe,
  Phys.\ Rev.\ D {\bf 73} (2006) 086003
  doi:10.1103/PhysRevD.73.086003
  [\arXiv{hep-th/0506118}].

\bibitem{Hamilton:2006fh}
  A.~Hamilton, D.~N.~Kabat, G.~Lifschytz and D.~A.~Lowe,
  Phys.\ Rev.\ D {\bf 75} (2007) 106001
   Erratum: [Phys.\ Rev.\ D {\bf 75} (2007) 129902]
  doi:10.1103/PhysRevD.75.106001, 10.1103/PhysRevD.75.129902
  [\arXiv{hep-th/0612053}].
  
\bibitem{Hamilton:2006az}
  A.~Hamilton, D.~N.~Kabat, G.~Lifschytz and D.~A.~Lowe,
  Phys.\ Rev.\ D {\bf 74} (2006) 066009
  doi:10.1103/PhysRevD.74.066009
  [\arXiv{hep-th/0606141}].

\bibitem{Hamilton:2007wj}
  A.~Hamilton, D.~N.~Kabat, G.~Lifschytz and D.~A.~Lowe,
  AMS/IP Stud.\ Adv.\ Math.\  {\bf 44} (2008) 85
  [\arXiv{arXiv:0710.4334} [hep-th]].

 
\bibitem{DiFrancesco:1997nk}
  P.~Di Francesco, P.~Mathieu and D.~Senechal,
  doi:10.1007/978-1-4612-2256-9
  
\bibitem{Aharony:1999ti}
  O.~Aharony, S.~S.~Gubser, J.~M.~Maldacena, H.~Ooguri and Y.~Oz,
  Phys.\ Rept.\  {\bf 323} (2000) 183
  doi:10.1016/S0370-1573(99)00083-6
  [\arXiv{hep-th/9905111}].
  
\bibitem{Almheiri:2014lwa}
  A.~Almheiri, X.~Dong and D.~Harlow,
  JHEP {\bf 1504} (2015) 163
  doi:10.1007/JHEP04(2015)163
  [\arXiv{arXiv:1411.7041} [hep-th]].
 
\bibitem{Swingle:2009bg}
  B.~Swingle,
  Phys.\ Rev.\ D {\bf 86} (2012) 065007
  doi:10.1103/PhysRevD.86.065007
  [\arXiv{arXiv:0905.1317} [cond-mat.str-el]].
\bibitem{Swingle:2012wq}
  B.~Swingle,
  \arXiv{arXiv:1209.3304} [hep-th].
\bibitem{Qi:2013caa}
  X.~L.~Qi,
  \arXiv{arXiv:1309.6282} [hep-th].

 
 
\bibitem{Pastawski:2015qua}
  F.~Pastawski, B.~Yoshida, D.~Harlow and J.~Preskill,
  JHEP {\bf 1506} (2015) 149
  doi:10.1007/JHEP06(2015)149
  [\arXiv{arXiv:1503.06237} [hep-th]].

\bibitem{Yang:2015uoa}
  Z.~Yang, P.~Hayden and X.~L.~Qi,
  JHEP {\bf 1601} (2016) 175
  doi:10.1007/JHEP01(2016)175
  [\arXiv{arXiv:1510.03784} [hep-th]].


\bibitem{Hayden:2016cfa}
  P.~Hayden, S.~Nezami, X.~L.~Qi, N.~Thomas, M.~Walter and Z.~Yang,
  JHEP {\bf 1611} (2016) 009
  doi:10.1007/JHEP11(2016)009
  [\arXiv{arXiv:1601.01694} [hep-th]].

\bibitem{Mintun:2015qda}
  E.~Mintun, J.~Polchinski and V.~Rosenhaus,
  Phys.\ Rev.\ Lett.\  {\bf 115} (2015) no.15,  151601
  doi:10.1103/PhysRevLett.115.151601
  [\arXiv{arXiv:1501.06577} [hep-th]].
  
\bibitem{deBoer:2016yiz}
  J.~de Boer, B.~Freivogel, L.~Kabir and S.~F.~Lokhande,
  JHEP {\bf 1707} (2017) 024
  doi:10.1007/JHEP07(2017)024
  [\arXiv{arXiv:1612.05265} [hep-th]].

\bibitem{Freivogel:2016zsb}
  B.~Freivogel, R.~A.~Jefferson and L.~Kabir,
  JHEP {\bf 1604} (2016) 119
  doi:10.1007/JHEP04(2016)119
  [\arXiv{arXiv:1602.04811} [hep-th]].

\bibitem{Dirac:1955uv}
  P.~A.~M.~Dirac,
  Can.\ J.\ Phys.\  {\bf 33} (1955) 650.
  doi:10.1139/p55-081

\bibitem{Donnelly:2015hta}
  W.~Donnelly and S.~B.~Giddings,
  Phys.\ Rev.\ D {\bf 93} (2016) no.2,  024030
   Erratum: [Phys.\ Rev.\ D {\bf 94} (2016) no.2,  029903]
  doi:10.1103/PhysRevD.94.029903, 10.1103/PhysRevD.93.024030
  [\arXiv{arXiv:1507.07921} [hep-th]].

\bibitem{Donnelly:2015taa}
  W.~Donnelly, D.~Marolf and E.~Mintun,
  Class.\ Quant.\ Grav.\  {\bf 33} (2016) no.2,  025010
  doi:10.1088/0264-9381/33/2/025010
  [\arXiv{arXiv:1510.00672} [hep-th]].

\bibitem{Miyaji:2015fia}
  M.~Miyaji, T.~Numasawa, N.~Shiba, T.~Takayanagi and K.~Watanabe,
  Phys.\ Rev.\ Lett.\  {\bf 115} (2015) no.17,  171602
  doi:10.1103/PhysRevLett.115.171602
  [\arXiv{arXiv:1506.01353} [hep-th]].
\bibitem{Nakayama:2015mva}
  Y.~Nakayama and H.~Ooguri,
  JHEP {\bf 1510} (2015) 114
  doi:10.1007/JHEP10(2015)114
  [\arXiv{arXiv:1507.04130} [hep-th]].
\bibitem{Nakayama:2016xvw}
  Y.~Nakayama and H.~Ooguri,
  JHEP {\bf 1610} (2016) 085
  doi:10.1007/JHEP10(2016)085
  [\arXiv{arXiv:1605.00334} [hep-th]].
\bibitem{Verlinde:2015qfa}
  H.~Verlinde,
  \arXiv{arXiv:1505.05069} [hep-th].
\bibitem{Lewkowycz:2016ukf}
  A.~Lewkowycz, G.~J.~Turiaci and H.~Verlinde,
  JHEP {\bf 1701} (2017) 004
  doi:10.1007/JHEP01(2017)004
  [\arXiv{arXiv:1608.08977} [hep-th]].
\bibitem{Hawking:1974sw}
  S.~W.~Hawking,
  Commun.\ Math.\ Phys.\  {\bf 43} (1975) 199
   Erratum: [Commun.\ Math.\ Phys.\  {\bf 46} (1976) 206].
  doi:10.1007/BF02345020
\bibitem{Bekenstein:1973ur}
  J.~D.~Bekenstein,
  Phys.\ Rev.\ D {\bf 7} (1973) 2333.
  doi:10.1103/PhysRevD.7.2333


\bibitem{Balasubramanian:2013lsa}
  V.~Balasubramanian, B.~D.~Chowdhury, B.~Czech, J.~de Boer and M.~P.~Heller,
  Phys.\ Rev.\ D {\bf 89} (2014) no.8,  086004
  doi:10.1103/PhysRevD.89.086004
  [\arXiv{arXiv:1310.4204} [hep-th]].
\bibitem{Headrick:2014eia}
  M.~Headrick, R.~C.~Myers and J.~Wien,
  JHEP {\bf 1410} (2014) 149
  doi:10.1007/JHEP10(2014)149
  [\arXiv{arXiv:1408.4770} [hep-th]].
\bibitem{Balasubramanian:2014sra}
  V.~Balasubramanian, B.~D.~Chowdhury, B.~Czech and J.~de Boer,
  JHEP {\bf 1501} (2015) 048
  doi:10.1007/JHEP01(2015)048
  [\arXiv{arXiv:1406.5859} [hep-th]].
\bibitem{Balasubramanian:2016xho}
  V.~Balasubramanian, A.~Bernamonti, B.~Craps, T.~De Jonckheere and F.~Galli,
  JHEP {\bf 1612} (2016) 094
  doi:10.1007/JHEP12(2016)094
  [\arXiv{arXiv:1609.03991} [hep-th]].
\bibitem{Faulkner:2017vdd}
  T.~Faulkner and A.~Lewkowycz,
  JHEP {\bf 1707} (2017) 151
  doi:10.1007/JHEP07(2017)151
  [\arXiv{arXiv:1704.05464} [hep-th]].
\bibitem{Engelhardt:2016wgb}
  N.~Engelhardt and G.~T.~Horowitz,
  Class.\ Quant.\ Grav.\  {\bf 34} (2017) no.1,  015004
  doi:10.1088/1361-6382/34/1/015004
  [\arXiv{arXiv:1605.01070} [hep-th]].
\bibitem{ElShowk:2011ag}
  S.~El-Showk and K.~Papadodimas,
  JHEP {\bf 1210} (2012) 106
  doi:10.1007/JHEP10(2012)106
  [\arXiv{arXiv:1101.4163} [hep-th]].
\bibitem{Kabat:2013wga}
  D.~Kabat and G.~Lifschytz,
  Phys.\ Rev.\ D {\bf 89} (2014) no.6,  066010
  doi:10.1103/PhysRevD.89.066010
  [\arXiv{arXiv:1311.3020} [hep-th]].
\bibitem{Hubeny:2012wa}
  V.~E.~Hubeny and M.~Rangamani,
  JHEP {\bf 1206} (2012) 114
  doi:10.1007/JHEP06(2012)114
  [\arXiv{arXiv:1204.1698} [hep-th]].
\bibitem{Headrick:2014cta}
  M.~Headrick, V.~E.~Hubeny, A.~Lawrence and M.~Rangamani,
  JHEP {\bf 1412} (2014) 162
  doi:10.1007/JHEP12(2014)162
  [\arXiv{arXiv:1408.6300} [hep-th]].
\bibitem{Ostlund:1995zz}
  S.~Ostlund and S.~Rommer,
  Phys.\ Rev.\ Lett.\  {\bf 75} (1995) 3537
  doi:10.1103/PhysRevLett.75.3537
  [\arXiv{cond-mat/9503107}].
\bibitem{Vidal:2003lvx}
  G.~Vidal,
  Phys.\ Rev.\ Lett.\  {\bf 93} (2004) no.4,  040502
  doi:10.1103/PhysRevLett.93.040502
  [\arXiv{quant-ph/0310089}].
\bibitem{Cirac:2004}
F.~Verstraete, J.I.~Cirac,
``Matrix product states represent ground states faithfully,"
 [\arXiv{arXiv:cond-mat/0505140} [cond-mat.str-el]].
\bibitem{Vidal:2007hda}
  G.~Vidal,
  Phys.\ Rev.\ Lett.\  {\bf 99} (2007) no.22,  220405
  doi:10.1103/PhysRevLett.99.220405
  [\arXiv{cond-mat/0512165}].
\bibitem{Peach:2017npp}
  A.~Peach and S.~F.~Ross,
  Class.\ Quant.\ Grav.\  {\bf 34} (2017) no.10,  105011
  doi:10.1088/1361-6382/aa6b0f
  [\arXiv{arXiv:1702.05984} [hep-th]].
\bibitem{Bhattacharyya:2016hbx}
  A.~Bhattacharyya, Z.~S.~Gao, L.~Y.~Hung and S.~N.~Liu,
  JHEP {\bf 1608} (2016) 086
  doi:10.1007/JHEP08(2016)086
  [\arXiv{arXiv:1606.00621} [hep-th]].
\bibitem{Blumenhagen:2009zz}
  R.~Blumenhagen and E.~Plauschinn,
  Lect.\ Notes Phys.\  {\bf 779} (2009) 1.
  doi:10.1007/978-3-642-00450-6

\bibitem{Balasubramanian:1998sn}
  V.~Balasubramanian, P.~Kraus and A.~E.~Lawrence,
  Phys.\ Rev.\ D {\bf 59} (1999) 046003
  doi:10.1103/PhysRevD.59.046003
  [\arXiv{hep-th/9805171}].

\bibitem{Polchinski:1995mt}
  J.~Polchinski,
  Phys.\ Rev.\ Lett.\  {\bf 75} (1995) 4724
  doi:10.1103/PhysRevLett.75.4724
  [\arXiv{hep-th/9510017}].

\bibitem{Freedman:1998tz}
  D.~Z.~Freedman, S.~D.~Mathur, A.~Matusis and L.~Rastelli,
  Nucl.\ Phys.\ B {\bf 546} (1999) 96
  doi:10.1016/S0550-3213(99)00053-X
  [\arXiv{hep-th/9804058}].

\bibitem{Parikh:2012kg}
  M.~Parikh and P.~Samantray,
  \arXiv{arXiv:1211.7370} [hep-th].


\end{thebibliography}\endgroup

\end{document}